\documentclass[aps,prd,twocolumn,superscriptaddress]{revtex4-2}
\usepackage{graphicx}
\usepackage{amsmath}
\usepackage{amssymb}
\usepackage{booktabs}
\usepackage{CJK}
\usepackage[colorlinks,citecolor=blue,linkcolor=blue,urlcolor=blue,anchorcolor=blue]{hyperref}
\allowdisplaybreaks

\begin{document}	
\begin{CJK*}{UTF8}{gbsn}
\title{Early dark energy triggered by spacetime dynamics that encodes \\ cosmic radiation-matter transition}
\author{Changcheng Jing (靖长成)}
\affiliation{Department of Astronomy, Beijing Normal University, Beijing 100875, China}
\author{Shuxun Tian (田树旬)}
\email[Corresponding author. ]{tshuxun@bnu.edu.cn}
\affiliation{Department of Astronomy, Beijing Normal University, Beijing 100875, China}
\author{Zong-Hong Zhu (朱宗宏)}
\email[Corresponding author. ]{zhuzh@bnu.edu.cn}
\affiliation{Department of Astronomy, Beijing Normal University, Beijing 100875, China}
\affiliation{School of Physics and Technology, Wuhan University, Wuhan 430072, China}
\date{\today}
\begin{abstract}
    Early dark energy (EDE),  introduced at the epoch of matter-radiation equality to alleviate the Hubble tension, has posed a new coincidence problem: why EDE appears at matter-radiation equality when their physics are completely unrelated? To solve this coincidence problem, we propose a new EDE model based on scalar-tensor gravity with the idea that EDE is triggered by spacetime dynamics that encodes cosmic radiation-matter transition. Our model can induce EDE naturally at matter-radiation equality without unnatural parameter tuning. Compared with other EDE models, a distinguishing feature of ours is that it can also induce a new energy component during cosmic matter-dark energy transition. This is testable with low-redshift observations.
\end{abstract}
\maketitle
\end{CJK*}
 
\section{\label{sec:Introduction}Introduction}
The tension between Hubble constant $H_{0}$ measured at low redshifts and that derived by fitting the $\Lambda$ cold dark matter ($\rm\Lambda$CDM) model to cosmic microwave background (CMB) data has acquired wildly attention \cite{Astrophys.J.874.4,PhysRevD.101.043533}. The best-fit of $\Lambda$CDM to the CMB is $H_{0}=67.36\pm 0.6\,\rm km/s/Mpc$ \cite{Astron.Astrophys.641.A6}. However, the measurement come from the local Cepheid calibrated SNe\,Ia observations, gives  $H_{0}=74.03\pm 1.42\,\rm km/s/ Mpc$ \cite{Astrophys.J.876.85}.  
Now the gap between those two measurements has reached $5\sigma$ \cite{Astrophys.J.Lett.934.L7}. As systematic errors in observations are unlikely to be responsible for such inconsistency \cite{PhysRevD.91.023518,Astrophys.J.802.20,Astrophys.J.818.132,Astrophys.J.867.108},
Hubble tension indicates the existence of new physics beyond the standard cosmology model ($\Lambda$CDM).
	
Early dark energy (EDE) is a possible solution for Hubble tension \cite{PhysRevLett.122.221301,PhysRevD.102.083513,PhysRevD.102.063527,PhysRevD.101.063523,PhysRevD.102.023523,PhysRevD.103.043528,PhysRevD.103.043529,DiValentino_2021,PhysRevD.106.063540}. Determination of $H_{0}$ with  CMB data requires fitting the integral expression $D^{*}_A=\frac{1}{H_{0}} \int_{0}^{z_{*}}\frac{d z}{E(z)}=r_{s}^*/\theta_{s}^*$, where $E(z)=H(z)/H_{0}$, and $z^{*}$ is the redshift when radiation decoupling,  $r_s$ is sound horizon, $\theta_{s}^*$ is precisely determined by CMB peak spacing,  $D^{*}_A$ is the angular diameter distance to last scattering. The sound horizon $r_s$ is given by
\begin{align}
	r_s=\int_{z_{*}}^{\infty}\frac{c_s }{H(z)} d z,
\end{align} 
where $c_s$ is the sound speed.
EDE increases $H$ near matter-radiation equality, resulting in a decrease of $r_s$ relative to the $\Lambda$CDM model. This leads a smaller $D^{*}_A$, which in turn results in a higher $H_0$ due to $H_{0}\propto {D_A^{*}}^{-1}$.

The widely studied canonical EDE model introduces the EDE component of Universe with a classical scalar field $\psi$ with a potential such as $V(\psi)=m^2(1-\cos[\psi/f])^n$, where $f$ is an energy scale \cite{PhysRevLett.113.251302,PhysRevLett.122.221301,DiValentino_2021}. Before the equality, the scalar field is frozen because of the Hubble friction term and acts as a cosmological constant. Around the equality, $\psi$ becomes dynamic and contributes a non-negligible energy component, i.e., the desired EDE. After that, the field dynamics become damped oscillation and its relative energy density can decay rapidly.

However, there is a coincidence problem in early dark energy approache, which was first recognized by \citet{PhysRevLett.124.161301}: why the EDE appears near matter-radiation equality when they does not have a direct physical connection? If EDE becomes active too early, it cannot provide sufficient modification on the sound horizon. If EDE decays too slowly after the equality, it would be disfavored by the data  \cite{PhysRevD.101.043533,PhysRevLett.124.161301}.
The time window in which EDE appears is very narrow and coincides with matter-radiation equality. 
For the canonical EDE model, this means that the energy scale $f$ must be fine-tuned.
Many works attempt to explain this coincidence. In this way, Refs.\cite{PhysRevLett.124.161301,JCAP.2021(06).063} construct a model that introduces EDE through nonminimal coupling of $\mathcal{O}(\rm eV)$ neutrinos and a scalar field. The key to solving the EDE coincidence problem is to find a suitable EDE trigger around matter-radiation equality. Other triggers include fluid equation of state (EOS) that encodes cosmic radiation-matter transition \cite{PhysRevD.103.043518,PhysRevD.107.103507} and the moment that dark matter being dominated \cite{PhysRevD.105.063535,PhysRevD.107.103523}.
	
This paper proposes a new EDE trigger --- spacetime dynamics that encodes cosmic radiation-matter transition. Inspired by the neutrino-assisted EDE model ($\nu$EDE) \cite{PhysRevLett.124.161301}, we also introduce a scalar field. Instead of introducing the coupling of the scalar field with neutrinos, we couple it with a geometric quantity $P$ that is a function of Ricci scalar $R$ and Gauss-Bonnet scalar $\mathcal{G}$ \cite{PhysRevD.74.046004}. The quantity $P$ equals zero in both radiation and matter-dominated era but nonzero at the transition. The geometry dynamics at the transition inject energy into the scalar field, which produces the desired evolution of EDE. The model can be formulated in the framework of the scalar-tensor theory of gravitation \cite{PhysRevD.61.023518,PhysRevD.63.063504,amendolatsujikawa2010}.
Compared with the canonical EDE model, our model eliminates the necessity of introducing a specialized energy scale to regulate the onset of EDE, thereby overcoming the coincidence associated with it. 

Another significant implication of our model is that it may lead to an additional energy component when the Universe transits from the matter-dominated epoch to the dark energy-dominated epoch. For a subclass within the model, it can generate an energy component persisting from around the equality to very low redshifts. This feature sets our model apart from other EDE models, as previous EDE models typically do not play a role in the cosmic evolution stages beyond the matter-radiation equality.

The structure of this paper is as follows. Section \ref{sec:Theory} presents our model. In Sec. \ref{sec:Dynamic}, we quantitatively study the evolution of the Universe with special parameters. In Sec. \ref{sec:late time}, we study the model in late-time Universe. Section \ref{sec:05} presents elementary observational constraints. Conclusions are presented
in Sec. \ref{sec:conclution}.
		
\section{\label{sec:Theory} The model}
In this study, our investigation spans from the radiation-dominated era to matter-radiation equality and extends to the present epoch. The Universe contains radiation, matter, cosmological constant, and scalar field $\phi$ associated with EDE. Our model is described by the action
\begin{align}
    S=&\frac{1}{2\kappa}\int {\rm d}x^4 \sqrt{-g} \left[ 
    R-2\Lambda/c^2-\nabla_{\alpha}\phi\nabla^{\alpha}\phi\right.\nonumber\\
    &\qquad\qquad\,\left.+\alpha P(R,\mathcal{G})\phi
    -\beta R \phi^2\right]+S_{\rm m},
    \label{eq:lagrangian} 
\end{align}
where $R$ represents Ricci scalar, $\mathcal{G}$ is Gauss-Bonnet scalar, $P(R,\mathcal{G})$ is a function of $R$ and $\mathcal{G}$, $\alpha$ and $\beta$ are dimensionless constants, $\kappa=8\pi G/c^4$, $c$ is the speed of light. The function $P$ needs to be constructed to be 0 in both radiation and matter-dominated periods and cannot be ignored during the cosmic radiation-matter transition.  
The role of the potential term $\beta R\phi^2$ is to pull the scalar field $\phi$ back to the zero point during the matter-dominated period~\footnote{In general relativity, $R=0$ in the radiation-dominated era, and $R\neq0$ in the matter-dominated era.}.  Additionally, we introduce the cosmological constant $\Lambda$ to account for the influence of dark energy in the late Universe.

A previously unsuccessful attempt is replacing $R\phi^2$ with the conventional potential $m^2\phi^2$ in Eq.~(\ref{eq:lagrangian}). In the model with $m^2\phi^2$, our calculations show that the time when EDE appears strongly depends on the value of $m$. Consequently, the model fails to address the EDE coincidence problem. By switching to $R \phi^2$, which links the mass of the scalar field to the spacetime geometry, this problem was resolved. The parameter $\beta$ does not require fine-tuning and typically assumes values of order unity, as demonstrated in Sec. \ref{sec:Dynamic}. Furthermore, with potential term $\beta R\phi^2$, EDE can decay faster than radiation, whereas normally, a scalar field with a quadratic potential can only decay as fast as the pressureless fluid \cite{PhysRevD.28.1243}. All the following discussions are based on the potential $R\phi^2$.

\subsection{Evolution equations}\label{sec:02A}
We can obtain the scalar field equation by varying the action (\ref{eq:lagrangian}) with respect to $\phi$, and the result is
\begin{align}
	\nabla_{\alpha}\nabla^{\alpha}\phi-\beta R \phi+\alpha P(R,\mathcal{G})/2=0.
	\label{eq:scalarfieldequation}
\end{align}
Varying the action (\ref{eq:lagrangian}) with respect to $g_{uv}$ gives the gravitational field equations
\begin{widetext}
\begin{align}
    &( R^{ab} -  \tfrac{1}{2} g^{ab} R) (4\alpha P_{G}{}\phi {}^{;c}{}_{;c}+8 \alpha \phi {}_{;c} P_{G}{}{}^{;c}+\beta\phi^2-1) + 2\alpha P_{R}{}{}^{;(a} \phi {}^{;b)} + \phi {}^{;a} \phi {}^{;b}(1  - 2 \beta) + \alpha P_{R}{} (\phi {}^{;a}{}^{;b} -  g^{ab} \phi {}^{;c}{}_{;c})\nonumber\\
    &+ 4 \alpha R\phi {}^{;(a} P_{G}{}{}^{;b)} - 8 \alpha R^{(a}{}_{c} \phi {}^{;b)} P_{G}{}{}^{;c} - 8 \alpha R^{(a}{}_{c}P_{G}{}{}^{;b)} \phi {}^{;c} +\tfrac{1}{2} g^{ab} \bigl[- \frac{2 \Lambda}{c^2} -  \phi {}_{;c} (4 \alpha P_{R}{}{}^{;c} + \phi {}^{;c} - 4 \beta \phi {}^{;c})\bigr]\nonumber\\
    & - 4 \alpha (R^{a}{}_{c}{}^{b}{}_{d} + R^{a}{}_{d}{}^{b}{}_{c}-8 \alpha g^{ab} R_{cd}) P_{G}{}{}^{;c} \phi {}^{;d}+P_{G}{} \bigl[ 2 \alpha R \phi {}^{;a}{}^{;b}  - 8 \alpha R^{(a}{}_{c} \phi {}^{;b)}{}^{;c} + 4 \alpha g^{ab} R_{cd} \phi {}^{;c}{}^{;d} - 4 \alpha R^{a}{}_{c}{}^{b}{}_{d} \phi {}^{;c}{}^{;d}\bigr]\nonumber\\
    &+\phi \left[- \alpha P_{R}{} R^{ab} + P_{G}{} (4 \alpha R^{ac} R^{b}{}_{c} - 2 \alpha R^{ab} R + 4 \alpha R^{cd} R^{a}{}_{c}{}^{b}{}_{d} - 2 \alpha R^{acde} R^{b}{}_{cde}) + \alpha P_{R}{}{}^{;a}{}^{;b} - 2 \beta \phi {}^{;a}{}^{;b} + 4 \alpha R^{ab} P_{G}{}{}^{;c}{}_{;c}\right.\nonumber\\
    &\left.+ 2 \alpha R (P_{G}{}{}^{;a}{}^{;b} -  g^{ab} P_{G}{}{}^{;c}{}_{;c}) + \tfrac{1}{2} g^{ab} (P \alpha - 2 \alpha P_{R}{}{}^{;c}{}_{;c} + 4 \beta \phi {}^{;c}{}_{;c} + 8 \alpha  R_{cd} P_{G}{}{}^{;c}{}^{;d}) - 8 \alpha R^{(a}{}_{c} P_{G}{}{}^{;b)}{}^{;c} - 4 \alpha  R^{a}{}_{d}{}^{b}{}_{c} P_{G}{}{}^{;c}{}^{;d}\right]\nonumber\\
    &+\kappa T^{ab}=0,
\end{align}
\end{widetext}
where the parenthesis in the upper and lower subscripts represents symmetry operator, e.g., $I^{(ab)}=(I^{ab}+I^{ba})/2$, and the 
semicolon represents covariant derivative for simplicity. This result is obtained based on the \texttt{xAct} package \footnote{\url{http://xact.es}}. 
In Eq. (\ref{eq:scalarfieldequation}), $P(R,\mathcal{G})$ acts as the source of scalar field $\phi$, which is similar to the trace of the neutrino energymomentum tensor in Ref. \cite{PhysRevLett.124.161301}. 
The function $P$ needs to be constructed to fulfill the following requirements: it equals zero during both radiation and matter-dominated era, and possesses a sufficiently large value during the transition. This will drive the scalar field away from the zero point, thereby induce EDE.

The Universe is assumed to be described by the flat Friedmann-Lema{\^\i}tre-Robertson-Walker (FLRW) metric.
Substituting $ds^2=-\gamma^2{\rm d}t^2+a^2{\rm d}\mathbf{x}^2$ ($\gamma$ and $a$ are functions of time $t$)  into action (\ref{eq:lagrangian}), variation with respect to $\gamma$ and $\phi$, replacing $\gamma$ with $c$ (choosing a gauge), we obtain the following equations of motion \footnote{The details are integrated into a public \texttt{Mathematica} code, which is available at GitHub (\url{https://github.com/JingChang-cheng/MG}).} 
\begin{align}
	&H^2-\frac{\dot{\phi}^2}{6}-\beta (\phi^2H^2+2  H \phi\dot{\phi})+\alpha H \dot{\phi} \bigg(\frac{4 H^2 P_{\mathcal{G}}}{c^4}+P_R\bigg)\nonumber\\
	&\quad+\alpha \phi
	\bigg[\frac{4 H^3 \dot{P}_{\mathcal{G}}- (H^2+\dot{H}) (4 H^2 P_{\mathcal{G}}+c^2 P_R)}{c^2}+\frac{Pc^2}{6} \nonumber\\
     &\quad + H \dot{P}_{R}\bigg]=\frac{8 \pi  G}{3}(\rho_{\rm r}+\rho_{\rm m}+\rho_\Lambda),
	\label{eq:Friedmannequation}\\
	&\ddot{\phi}+3 H \dot\phi+6 \beta  \phi (2 H^2+\dot{H})=\alpha Pc^2/2\label{eq:phi},
\end{align}
where  $\dot{}\equiv{\rm d}/{\rm d}t$, $H\equiv\dot{a}/{a}$, $P_{\mathcal{G}}\equiv\partial P/\partial\mathcal{G}$, $P_{R}\equiv\partial P/\partial R$, the subscripts $\{{\rm r,m},\Lambda\}$ denotes radiation, pressureless matter and the cosmological constant, respectively. Note that $\rho_\Lambda=\Lambda/8\pi G$ in our conventions. We also obtain the second Friedmann equation by varying the action (\ref{eq:lagrangian}) with respect to $a$. We checked that this equation is not independent of Eqs. (\ref{eq:Friedmannequation}) and (\ref{eq:phi}). The right side of Eq. (\ref{eq:Friedmannequation}) corresponds to the current standard model of cosmology \cite{Astron.Astrophys.641.A6}. To characterize the dynamics of EDE and the Universe, we define the relative energy density $\Omega_i=8\pi G\rho_i/(3H^2)$, where $i\in\{{\rm r,m},\Lambda,{\rm EDE}\}$ and the effective quantities for EDE could be directly read from Eq.~(\ref{eq:Friedmannequation}). Our model goes back to standard $\Lambda$CDM when $\phi\to0$.

We require $\Omega_{\rm EDE}\approx10\%\ll1$ during the cosmic radiation-matter transition. Consequently, the coupling between the scalar field $\phi$ and the geometric quantity $P$ should be weak ($\alpha\ll1$). This enables us to employ a perturbation method to solve Eqs. (\ref{eq:Friedmannequation}) and (\ref{eq:phi}). Specifically, we consider $\alpha$ as a small quantity of first order and define $H=H_{\rm std}+H_{\rm EDE}$. In this expression, the first term corresponds to the background value of $H$ derived from the standard $\Lambda$CDM model, while the second term represents the correction induced by the EDE field. 
In the perturbation framework, Eq.  (\ref{eq:phi}) indicates that both $\alpha$ and $\phi$ are small quantities of the same order. Consequently, we need only consider the background value $H_{\rm std}$ when solving Eq. (\ref{eq:phi}) to give $\phi$. Furthermore, Eq. (\ref{eq:Friedmannequation}) defines $\Omega_{\rm EDE}$, whose value can be approximately calculated based on $\phi$ and $H_{\rm std}$. The difference between this and the exact solution appears at higher order and hence is small.

\subsection{Choice of function $P$}
For our purpose, $P$ should be equal to zero in both the radiation and matter-dominated era and become nonzero around the matter-radiation equality. On the other hand, to avoid the fine-tuning problem, we aim for $P$ to be of the same order as $R$. We choose to contruct $P$ with Ricci scalar $R$ and Gauss-Bonnet scalar  $\mathcal{G}$, where 
$R=6(2H^2+\dot{H})/c^2=3H^2(1-3w)/c^2$ and $\mathcal{G}=24H^2(H^2+\dot{H})/c^4=-12H^4 (1+3w)/c^4 $ in general relativity,
with $w$ representing the EOS of perfect fluid. 
It is shown that $R=0$ for $w=1/3$ (radiation era), and $4R^2+3\mathcal{G}=0$ for $w=0$ (matter era). Therefore, $P$ can be chosen as
\begin{align}
	P(R,\mathcal{G})\equiv\Big(\frac{4R^2+3\mathcal{G}}{\mathcal{G}}\Big)^n R,\label{Pv}
\end{align}
where $n$ is a positive integer. The denominator $\mathcal{G}$ is introduced to ensure that $P$ has the same dimension with $R$. 
\begin{figure}[!b]
	\centering
	\includegraphics[width=0.48\textwidth]{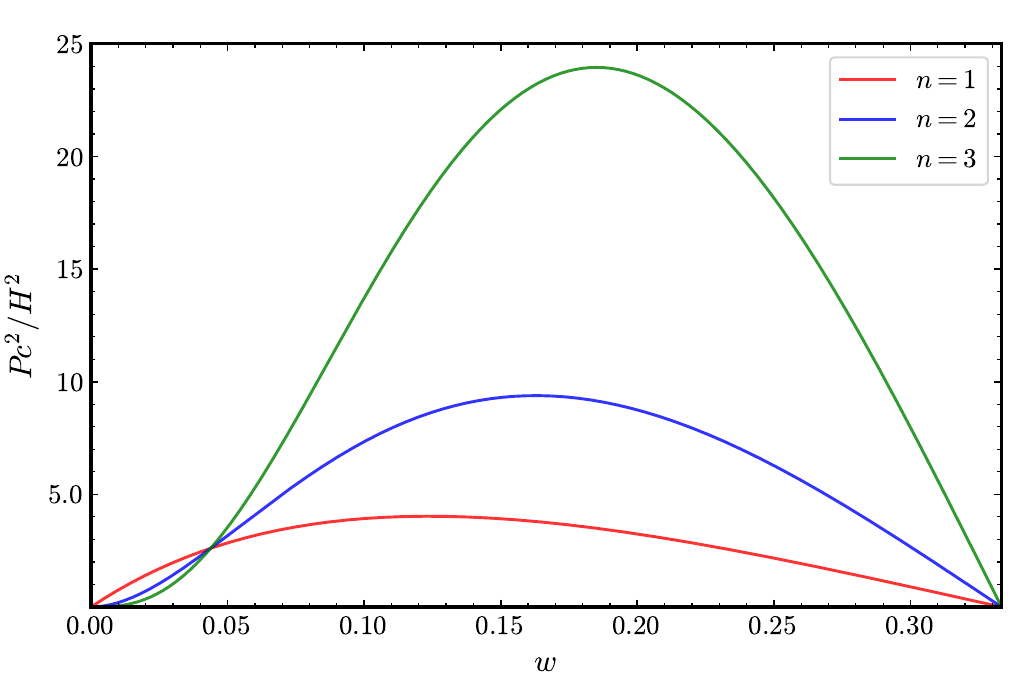}
        \caption{ The quantity $Pc^2/H^2$ is examined as a function of the equation of state (EOS) parameter $w$ within the context of the $\Lambda$CDM model. The red line is for $n=1$, the blue line is for $n=2$, and the green line is for $n=3$. The parameter $n$ is crucial in our model: the peak of $\Omega_{\rm EDE}$ appears earlier for a larger $n$.
        }
	\label{fig:f}
\end{figure}
In general relativity, we show the value of $Pc^2/{H}^2$ as a function of $w$ for different $n$ in Fig. \ref{fig:f}. As expected, it equals $0$ at $w=0$ and $1/3$. With an increasing value of $n$, the peak of the function $Pc^2/{H}^2$ moves to $w=1/3$ (radiation-dominated period), and the corresponding peak of the EDE relative density $\Omega_{\rm EDE}$ appears earlier (see Sec. \ref{sec:Dynamic}).
	
However, when considering the late-time Universe, $\mathcal{G}$ transports from a negative value to a positive value as our Universe shifts from a matter-dominated epoch to a dark energy-dominated one.  This introduces a singularity in the $P(R,\mathcal{G})$ function.   To circumvent this issue, we modify the function $P(R,\mathcal{G})$ as follows:
\begin{align}
	P(R,\mathcal{G})\equiv\left[ \frac{4R^2+3\mathcal{G}}{\mathcal{G}-(\gamma\Lambda/c^2)^2}\right] ^n R,\label{PvL}
\end{align}
where $\gamma$ is a dimensionless constant. When cosmological constant $\Lambda$ is negligible, Eq. (\ref{PvL}) backs to  (\ref{Pv}). A nonzero value of $\gamma$ should be chosen to ensure the absence of the singularity. Calculating the values of $\mathcal{G}$ and $P$ for the de Sitter Universe gives $\gamma>\sqrt{3}$. This modification does not impact our previous discussion concerning EDE because during the matter-radiation equality, $\mathcal{G}\sim10^{19}(\Lambda/c^2)^2$, where $(\gamma\Lambda/c^2)^2$ is negligible as long as $\gamma$ is not excessively large.
  
\section{\label{sec:Dynamic}EDE dynamics}
The initial conditions of the scalar field at deep radiation era are assumed to be $\phi=0$ and $\dot{\phi}=0$. In the radiation-dominated era, the evolution of $\phi$ is dominated by Hubble friction due to $R=0$. Consequently, $\phi$ stays around $0$ throughout the radiation era. 
Then, in the process of radiation-matter transition, the scalar field $\phi$ is kicked away from the origin by the geometric term $P$ and leads to the emergence of EDE. This process is similar to the $\nu$EDE model when neutrinos become non-relativistic \cite{PhysRevLett.124.161301}, in which a similar scalar field kicked away from its initial position by a term related to the massive neutrinos.
Subsequently, during the matter-dominated epoch, Eq. (\ref{eq:phi}) transforms into a damped oscillatory equation, causing $\phi$ to evolve towards zero. 

In this section, we study our theory with different parameters. Here, we set the cosmological constant $\Lambda$ to 0 because it was negligible in the early universe.  A successful EDE must become active shortly before matter-radiation equality and rapidly decay soon after \cite{PhysRevLett.124.161301}.
Numerical results show our theory can match those requirements with suitable parameters.

\subsection{Evolution of $\Omega_{\rm EDE}$ at matter-dominated epoch}
A successful EDE model should decay soon after matter-radiation equality \cite{PhysRevLett.124.161301}. Models proposed in Refs. \cite{PhysRevD.94.103523,ALEXANDER2019134830} achieve a decay of EDE as fast as or faster than radiation, but suffer from the coincidence problem. Here, we show that our model can rapidly decay in the matter-dominated epoch. For this case, utilizing Eqs. (\ref{eq:Friedmannequation}), (\ref{eq:phi}) and $w=0$, we attain 
\begin{equation}
	{\phi}''+ \frac{3}{2} {\phi}'+3 \beta\phi
	=0,\label{eq:phi1matter}
\end{equation}
and
\begin{equation}\label{eq:valueofOmega}
  \Omega_{\rm EDE} = \left\{
  \begin{array}{lll}
    \beta\phi^2+(\phi')^2/6 +2\beta\phi\phi', & & n\geq2, \\
                                              & &         \\
    \beta\phi^2+(\phi')^2/6+2\beta\phi\phi'   & &         \\[1.5mm]
    \quad\ \ +\alpha (27\phi'+18\phi)/2,      & & n=1,
  \end{array}
  \right.
\end{equation}
where  $\prime\equiv{\rm d}/{\rm d}N$, $N\equiv\ln a$.
The motion equation of  ${\phi}$ transforms into a damped oscillation equation \cite{Landanvolume1}, which have an analytical solution
\begin{align}
	{\phi}(N)=e^{-3N/4}(A_1 e^{\omega^{+}N}+A_2 e^{-\omega^{+}N}),\label{eq:solutionphi1matter}
\end{align}
where $A_1$ and $A_2$ are constants, and $\omega^{\pm}\equiv\sqrt{\pm9\mp48\beta}/4$.
	
For $n\geq2$,
if $\beta<3/16$, the last term on the right hand of Eq. (\ref{eq:solutionphi1matter}) decay faster than the first one, so ${\phi}(N)= A_1 \exp[ (4 \omega^{+}-3)N/4] $. Then we can attain $\Omega_{\rm EDE}\propto  \exp[  \left(4\omega^{+}-3\right) N/2] =a^{(4\omega^{+}-3)/2}$. On the other hand, we know the relative energy density $\Omega_{\rm r}=\rho_{\rm r}/\rho_{\rm m}\propto a^{-1}$, so the requirement that EDE decay faster than radiation is satisfied for $\beta>1/6$ in this case. 
If $\beta=3/16$, $\Omega_{\rm EDE}\propto a^{-3/2}$,  at which point ${\phi}$ is critically damped. 
If $\beta>3/16$, $A_1=A_2$ must be satisfied in order to guarantee that the solution is meaningful, which leads ${\phi}(N)=A e^{-3N/4}\cos(\omega^{-}N)$, where $A=2A_1$.  Together with Eq. (\ref{eq:valueofOmega}), we can attain
\begin{align}
    \Omega_{\rm EDE}&=\frac{A^2}{32}e^{-3 N/2}\left[ (3-16 \beta ) \cos (2\omega^{-} N)\right.\nonumber\\
    &\quad\left.+(1-8 \beta ) 4\omega^{-} \sin (2\omega^{-} N)\right] ,
\end{align}
which is oscillating and decaying. It is decay as $a^{-3/2}$, that is faster than radiation too.
	
For $n=1$, since scalar field ${\phi}$ decay exponentially, $\Omega_{\rm EDE}= \alpha \left[27 {\phi}'+18 {\phi}\right]/2$. If $\beta<3/16$, $\Omega_{\rm EDE}\propto \exp[ (4\omega^{+}-3) N/4] =a^{ (4\omega^{+}-3)/4}$, which can't decay as fast as radiation.  If $\beta\geq3/16$, $\Omega_{\rm EDE}$ decay as $a^{-3/4}$, which is slower than radiation. 

In summary, the viable parameter space that can achieve a faster decay of EDE than radiation during the matter-dominated period is $\{n\geq2,\beta>1/6\}$.

\subsection{Numerical solution for the transition}
This section aims to demonstrate that our model can indeed achieve the desired evolution of EDE, which appears near the matter-radiation equality and decays rapidly. Specifically, in the perturbation framework (see the discussions at the end of Sec. \ref{sec:02A}), for a given $H_{\rm std}$, we numerically solve Eq. (\ref{eq:phi}) to obtain $\phi$, and then calculate $\Omega_{\rm EDE}$ based on the solution of $\phi$ and $H_{\rm std}$ with Eq. (\ref{eq:Friedmannequation}). We set the initial conditions at the deep radiation era, and the scalar field evolves from $\phi=0$ and $\dot{\phi}=0$. The density of matter $\rho_{\rm m}$ and radiation $\rho_{\rm r}$ satisfies $\rho_{\rm m}=\rho_{\rm m0}e^{-3N}$, $\rho_{\rm r}=\rho_{\rm r0}e^{-4N}$, where  $\rho_{\rm m0}$ and $\rho_{\rm r0}$ represent the corresponding values today.

\begin{figure*}[t]
    \centering
    \includegraphics[width=0.48\textwidth]{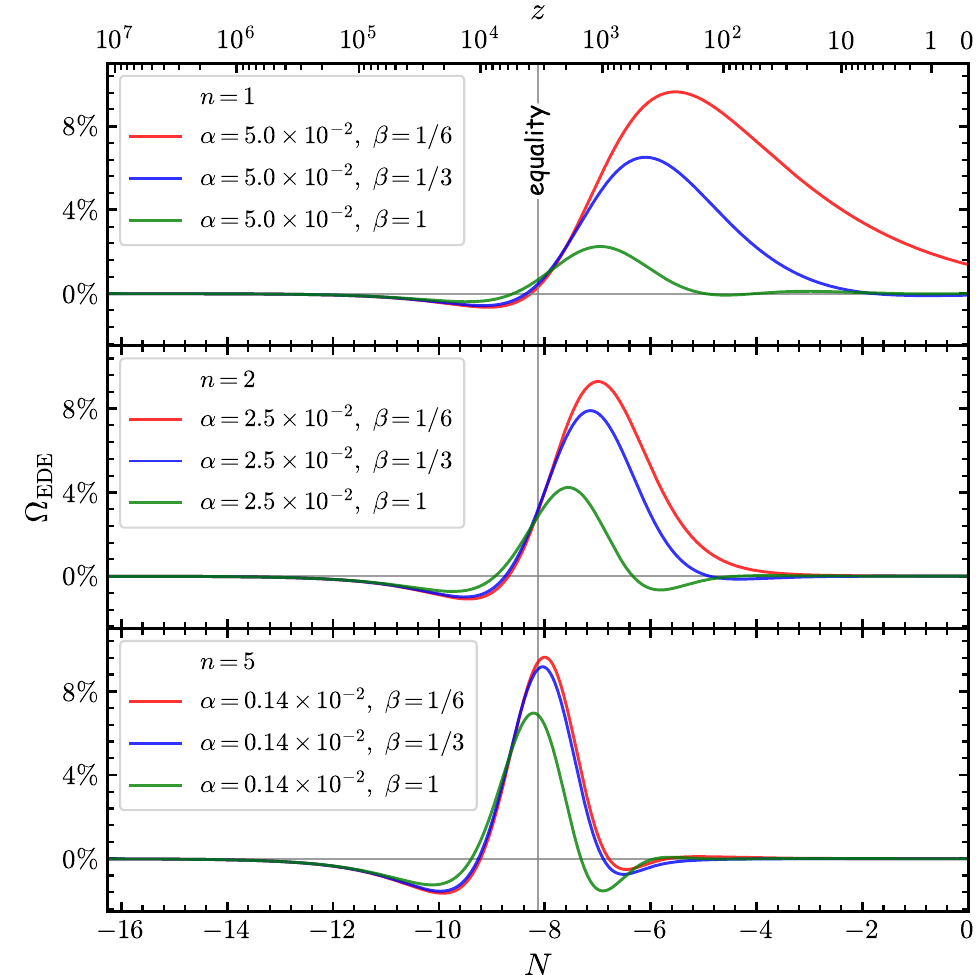}
    \includegraphics[width=0.48\textwidth]{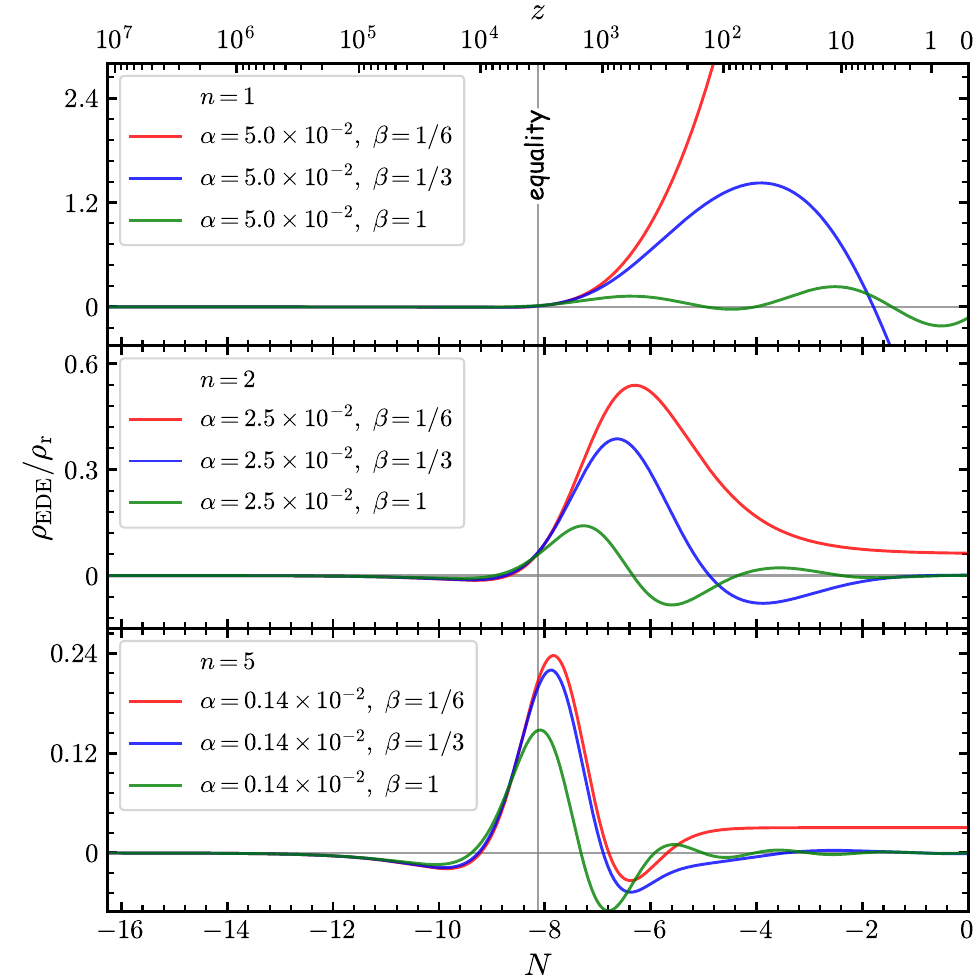}
    \caption{Evolution of $\Omega_{\rm EDE}$ (left) and $\rho_{\rm EDE}/\rho_{\rm r}$ (right) as a function of $N$ and redshift $z$ for different parameters.}
    \label{fig:EvolutionOmegaede}
\end{figure*}

The left panel of Fig. \ref{fig:EvolutionOmegaede} shows the evolution of EDE relative energy density $\Omega_{\rm EDE}$. For $n=1$, since the peak appears much late to matter-radiation equality, the field cannot provide sufficient modification to the sound horizon formed before last scatting \cite{Astrophys.J.874.4,PhysRevD.101.043533}. Therefore, this case can not solve the Hubble tension. 

For the cases $n=2$ and $5$, the peaks appear near matter-radiation equality. Increasing $n$ can make the peak of $\Omega_{\rm EDE}$ move to left. On the other hand, EDE can decay rapidly after the equality. For $n=5$, $\alpha=0.14\times10^{-2}$ and $\beta=1/3$, $\Omega_{\rm EDE}$ decays to $0.05\%$  at last scattering. 
Different from Ref. \cite{PhysRevLett.124.161301}, there appear negative values in $\Omega_{\rm EDE}$, which is normal in modified gravity theory \cite{amendolatsujikawa2010}. In our model, the energy density of EDE is an effective energy density composed of specific geometric quantities and the scalar field $\phi$. Consequently, it allows for the possibility of a negative value of $\Omega_{\rm EDE}$. Negative $\Omega_{\rm EDE}$ indicates that the Hubble parameter $H$ is smaller than its counterpart in the $\Lambda$CDM model. In this scenario, the comoving sound horizon increases faster than the standard $\Lambda$CDM model. Due to this characteristic, our model requires a higher peak value of $\Omega_{\rm EDE}$ compared to other EDE models.
	
To compare the decay rate of EDE and radiation in the matter-dominated era, we plot $\rho_{\rm EDE}/\rho_{\rm r}$ as a function of $N$ in the right panel of Fig.  \ref{fig:EvolutionOmegaede}. It exhibits different properties according to different $\beta$ and $n$.
For $n=1$, as we have discussed before, EDE decays slower than radiation.  
For $n\geq2$, EDE can decay as fast as or faster than radiation after the peak. When $\beta=1/6$, $\rho_{\rm EDE}/\rho_{\rm r}$ evolves towards a nonzero constant. When $\beta=1/3 $ or $1$, $\rho_{\rm EDE}/\rho_{\rm r}$ oscillates and decays as the redshifts decrease, which is consistent with our expectation, indicating that $\phi$ evolves in an under-damped oscillation and that EDE decays faster than radiation. 

	
\section{\label{sec:late time}Low-redshifts dynamics}
Our model uses spacetime dynamics to trigger EDE. Similar to the radiation-matter transition, the matter-dark energy transition in the late Universe may also trigger a new energy component. This property is an important feature that distinguishes our model from other EDE models with different triggers (see Sec. \ref{sec:Introduction} for the discussions about \cite{PhysRevLett.124.161301,JCAP.2021(06).063,PhysRevD.103.043518,PhysRevD.107.103507,PhysRevD.105.063535,PhysRevD.107.103523}). In this section, we study the cosmic evolution at low redshifts (compared with matter-radiation equality). Dark energy is assumed to be described by the cosmological constant $\Lambda$ for simplicity, and we adopt Eq. (\ref{PvL}) for the function $P$. Note that the $\Lambda$ that appears in Eq. (\ref{PvL}) is used to avoid the denominator being equal to zero. 

\begin{figure*}[t]
    \includegraphics[width=0.45\textwidth]{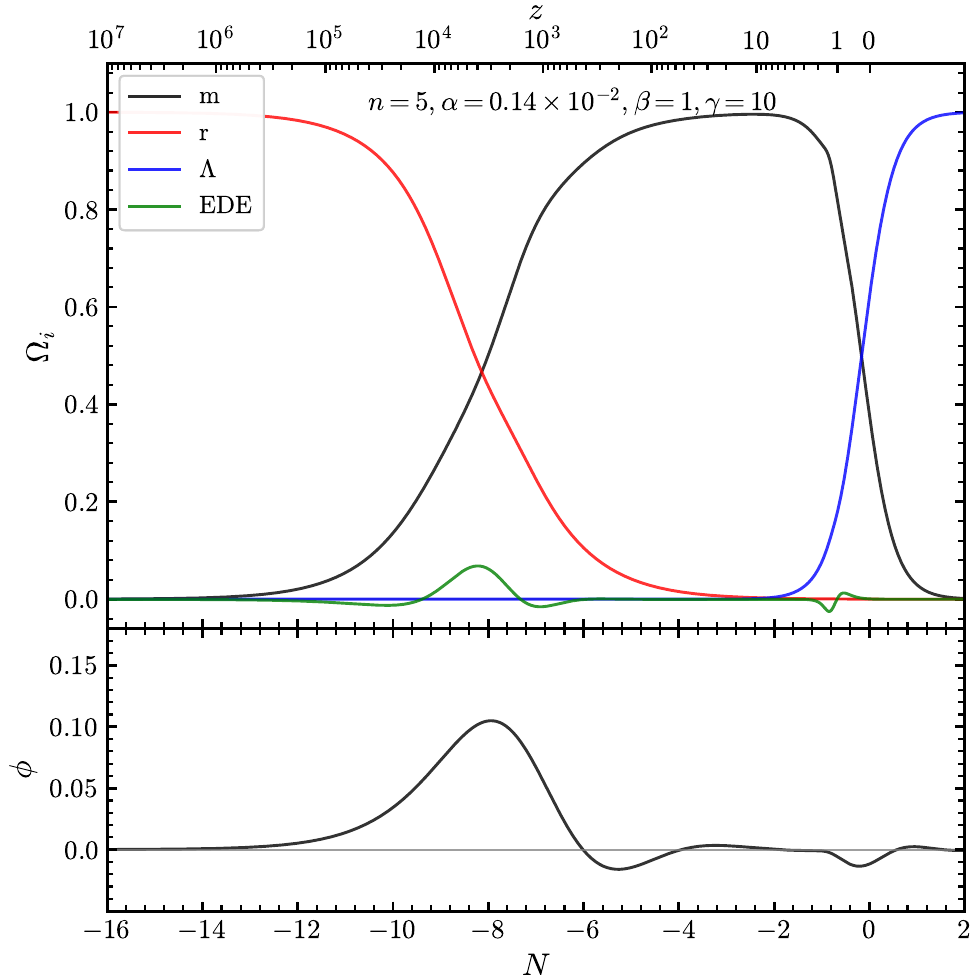}
    \includegraphics[width=0.45\textwidth]{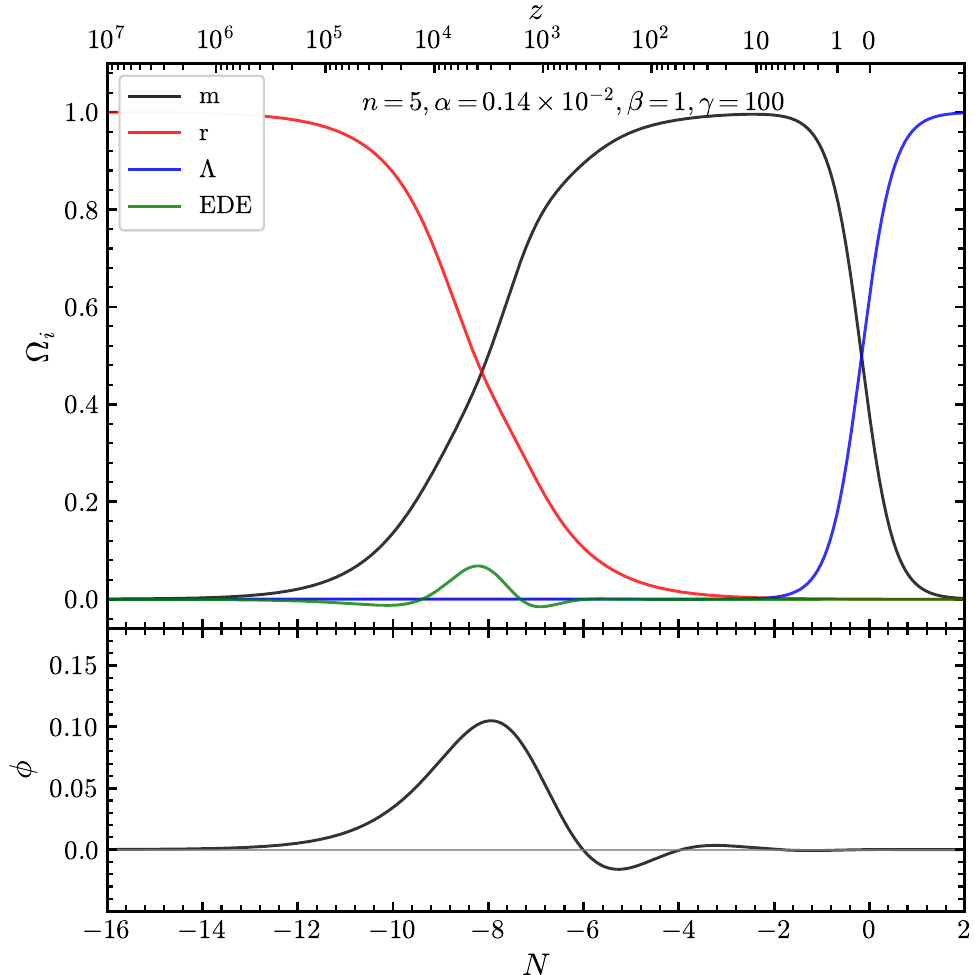}\\
    \includegraphics[width=0.45\textwidth]{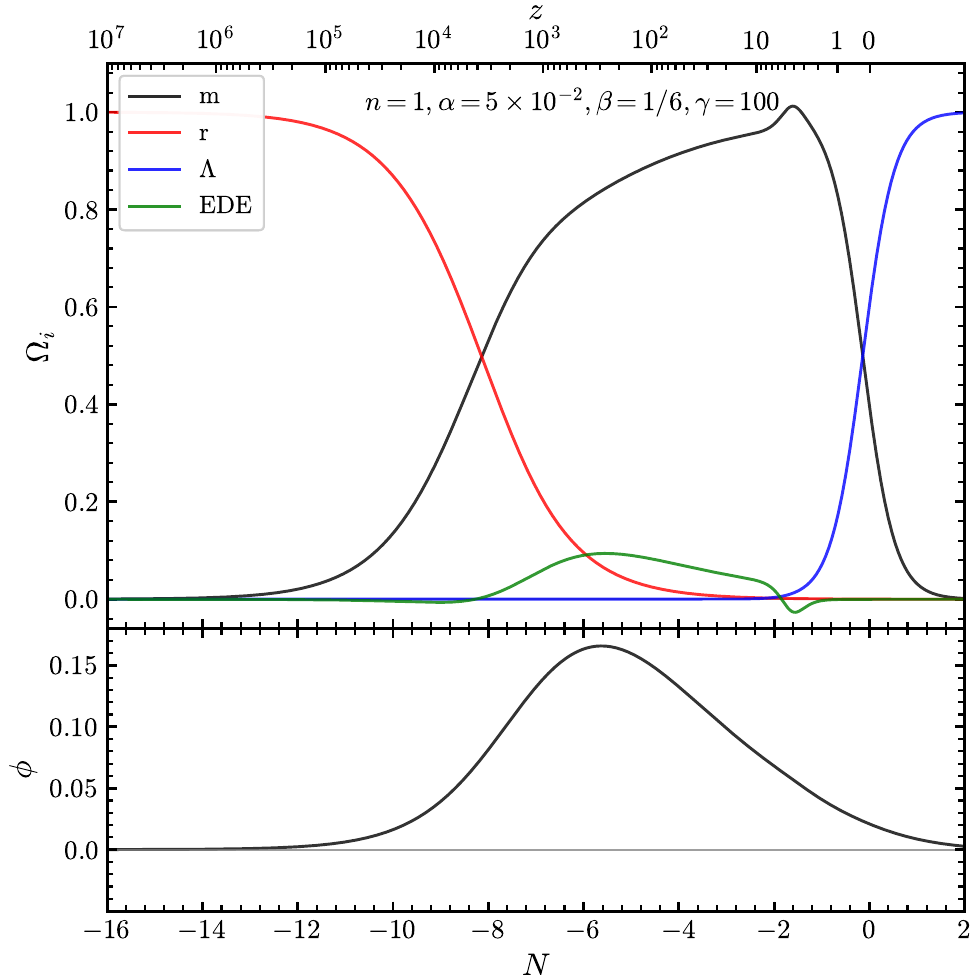}
    \includegraphics[width=0.45\textwidth]{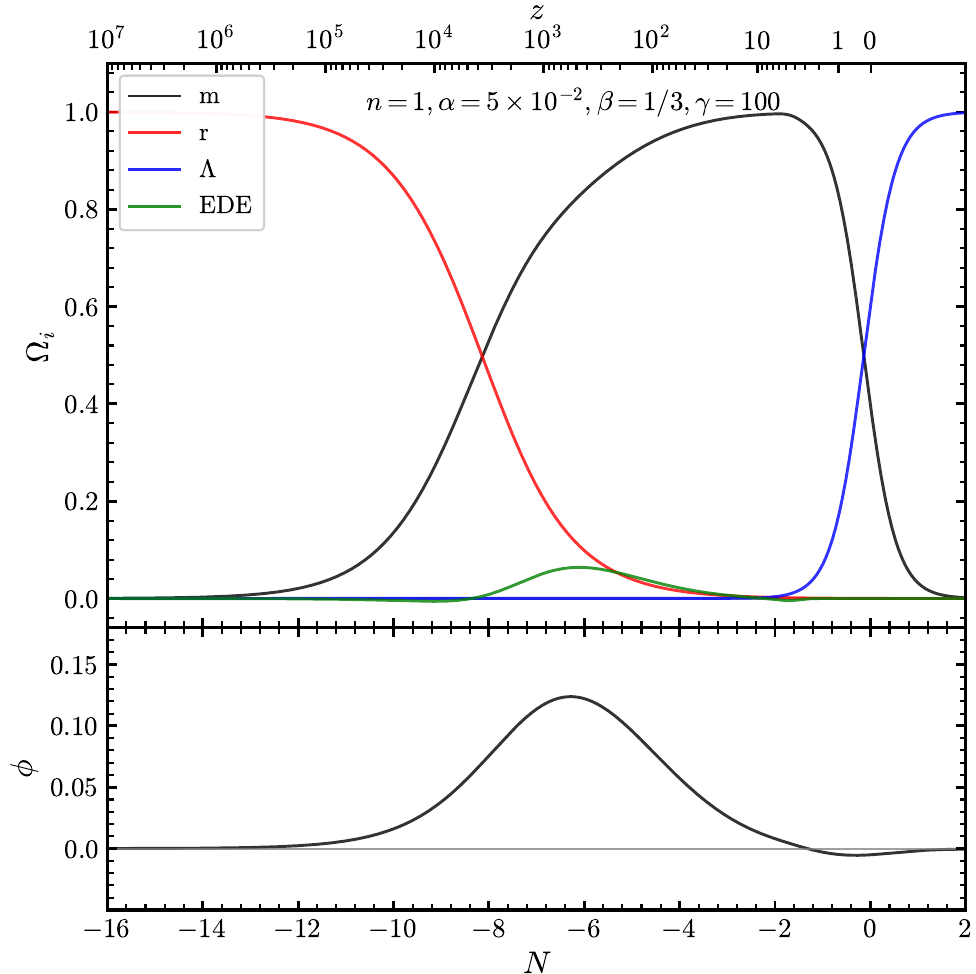}\\
    \caption{Evolution of matter (black), radiation (red), EDE (blue) and scalar field $\phi$ (black). Model parameters can be found in the figure.}
    \label{fig:EvolutionAll}
\end{figure*}

When dark energy starts to be dominated, $P$ deviates from zero, and $\phi$  is driven away from the zero point. This leads to a new energy component in the late-time Universe. Figure \ref{fig:EvolutionAll} show the evolution of $\Omega_{\rm EDE}$, $\Omega_{\rm \Lambda}$, $\Omega_{\rm m}$, $\Omega_{\rm r}$, and $\phi$ with $n=5$, $\alpha=0.14\times10^{-2}$, $\beta=1$, $\gamma=10$ (upper left) and $100$ (upper right). For the case of $\gamma=10$, $\phi$ was obviously triggered by not only radiation-matter transition but also matter-dark energy transition. The case of $\gamma=100$ shows that a large $\gamma$ can suppress the late-time dynamics of $\phi$ and result in a negligible $\Omega_\phi$. Current observations hint at the possibility of dynamical dark energy \cite{Zhao2017.NatAstron.1.627,Wang2022.PRD.106.063515}, which may be interpreted as the transition-induced energy component in our model. In addition, as we expected, comparing the two plots confirms that $\gamma$ does not affect the EDE dynamics around the matter-radiation equality.
	
Our model may be able to generate rich properties beyond the cosmic transitions. The bottom part of Fig. \ref{fig:EvolutionAll} shows the cosmic evolution for $n=1$. In the case of $\beta=1/6$ (left), the EDE is considerable from mater-radiation equality to low redshifts. If we take a slightly larger $\beta$, we can make EDE disappear completely at a higher redshift. For example, if we take $\beta=1/3$, EDE will disappear earlier ($z\approx10$, see the bottom right part of Fig. \ref{fig:EvolutionAll}). Therefore, in the case of $n = 1$, although our theory can not resolve the Hubble tension, it can produce a long-lasting dark energy component during the dark ages of the Universe. An extra energy component beyond the $\Lambda$CDM model is allowed to appear in the dark ages of the Universe. This possibility has been discussed by \cite{PLB.594.17.2004,PhysRevD.87.083009}, and our model with $n=1$ provides a theoretical realization.

\section{Observational constraints}\label{sec:05}
In this section, we perform elementary parameter constraints related only to cosmic background evolution. 

\subsection{Constraints on EDE}
Constraints on EDE require Cosmic Microwave Background (CMB) data. Given the absence of the evolution equation for cosmological perturbations, we could not utilize the complete CMB power spectrum for parameter fitting. Instead, we computed the CMB shift parameters \cite{PhysRevD.76.103533,PhysRevD.88.043522}
\begin{align}
R\equiv\sqrt{\Omega_mH_{0}^2} r^{*},\ l_a\equiv\pi r^*/r_{s}^*,
\end{align}
and employed them, along with $\Omega_b h^2$, for testing cosmological parameters, where $r^*$ represent the comoving distance of CMB. Ref. \cite{JCAP07(2019)005} generated the data points and covariance matrix for $[R, l_a, \Omega_b h^2]$ by fitting Planck 2018 results. Additionally, we incorporated data from the SH0ES data \cite{APJL.934.L7(2022)} and Pantheon plus data \cite{APJ.938.110(2022)}.

To reconcile our model with cosmological observations, we performed Markov Chain Monte Carlo (MCMC) analysis using the publicly available code \texttt{CosmoSIS} \cite{ZUNTZ201545}.  We adopted flat priors for the cosmological parameters {$\log\alpha$, $\beta$, $h$, $\Omega_m$, $\Omega_b h^2$} in the range of $(-3.5,-1)$, $(0,1)$, $(0.6, 0.8)$, $(0.2,0.4)$, and $(0.021,0.023)$, respectively. In addition to our model parameters, here $\Omega_m$ and $\Omega_b$ denote the present-day relative energy densities of matter and baryons respectively, and $h$ represents the Hubble constant in units of 100\,km/s/Mpc.
Additionally, We fixed the value $\gamma=100$ because it has little effect on the acoustic horizon formed in the early Universe. Separate MCMC analyses were conducted to compare models with $n=\{2,3,4,5\}$.

\begin{figure}[!t]
    \includegraphics[width=0.45\textwidth]{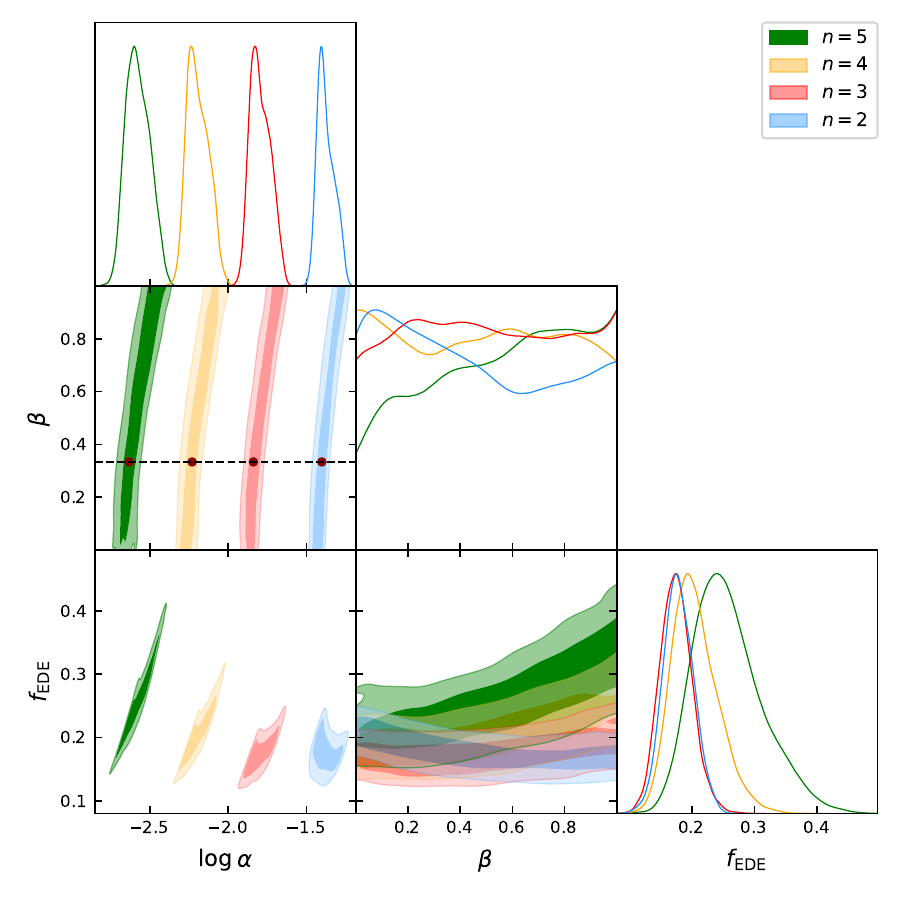}
    \caption{The marginalized 1D and 2D posterior distribution of $\log\alpha$, $\beta$ and $f_{\rm EDE}$ (peak value of $\Omega_{\rm EDE}$) for the EDE model with $n=5,4,3,2$. We highlight the best-fit parameter points with fixed values of $\beta=1/3$ and $\Omega_b h^2=0.022383$ in the $\log\alpha-\beta$ panel.}
    \label{fig:MCMC}
\end{figure}

\begin{figure}[!t]
    \includegraphics[width=0.45\textwidth]{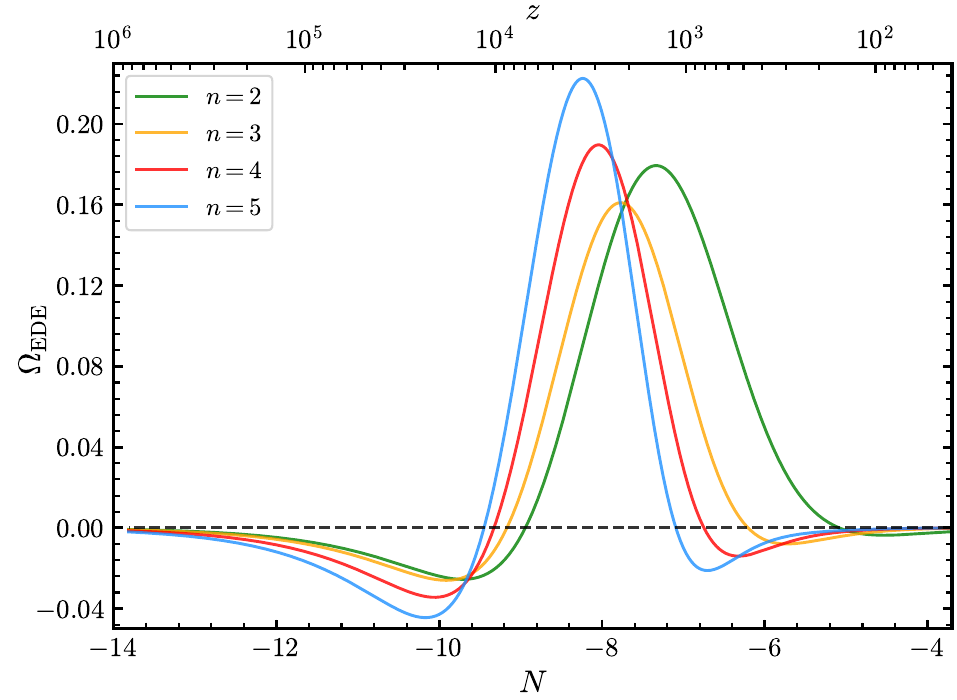}
    \caption{Evolution of $\Omega_{\rm EDE}$ for the marked parameter points in Fig. \ref{fig:MCMC}.}
    \label{fig:evolution}
\end{figure}

In Fig. \ref{fig:MCMC}, we show the marginalized 1D and 2D posterior distribution of $\log{\alpha}$, $\beta$, and $f_{\rm EDE}$ in our EDE model with $n=2,3,4,5$, where $f_{\rm EDE}$ represent the peak value of $\Omega_{\rm EDE}$. Note that $f_{\rm EDE}$ is not a parameter but a derived quantity of our model.  As we have discussed in Sec. \ref{sec:Dynamic}, the peak value of our EDE model should be higher than the previous model to solve the Hubble tension (see the marginalized 1D posterior distribution of $f_{\rm EDE}$ in Fig. \ref{fig:MCMC}).   In addition, Fig. \ref{fig:MCMC} shows that CMB shift parameters and SNe\,Ia data have weak constraints on $\beta$. The best-fit points with $\beta=1/3$ and $\Omega_b h^2=0.022383$ are marked in the $\alpha-\beta$ counter diagram, and the corresponding evolution of $\Omega_{\rm EDE}$ is plotted in Fig. \ref{fig:evolution}. 

One thing should be emphasized here. In our fitting, there are three parameters $\{\alpha,\beta,\Omega_bh^2\}$ that are strongly correlated with the CMB shift parameters and almost unrelated to the low-redshift SN\,Ia data. Meanwhile, CMB has only 3 data points and supernova has more than 1000 data points. Therefore, our EDE model may be overfitting to CMB but does not contribute to the posterior distribution of $H_0$. In other words, the posterior $H_0$ comes almost entirely from supernovae. In view of this, our result presented in Fig. \ref{fig:MCMC} does not mean a strong evidence for the existence of EDE even though $f_{\rm EDE}>0$ in high confidence level. Instead, Fig. \ref{fig:MCMC} demonstrates the ability of our model to decrease the acoustic horizon formed in the early Universe, which is widely used to explain how EDE solves the Hubble tension (see Sec. \ref{sec:Introduction} and references therein). In the future, we need global parameter constraints including the CMB angular power spectrum to determine whether our EDE model gives a better fit than the standard $\Lambda$CDM model, and whether our model can actually solve the Hubble tension.


\subsection{Constraints on the induced dynamical dark energy}
\begin{figure}[!t]
    \includegraphics[width=0.45\textwidth]{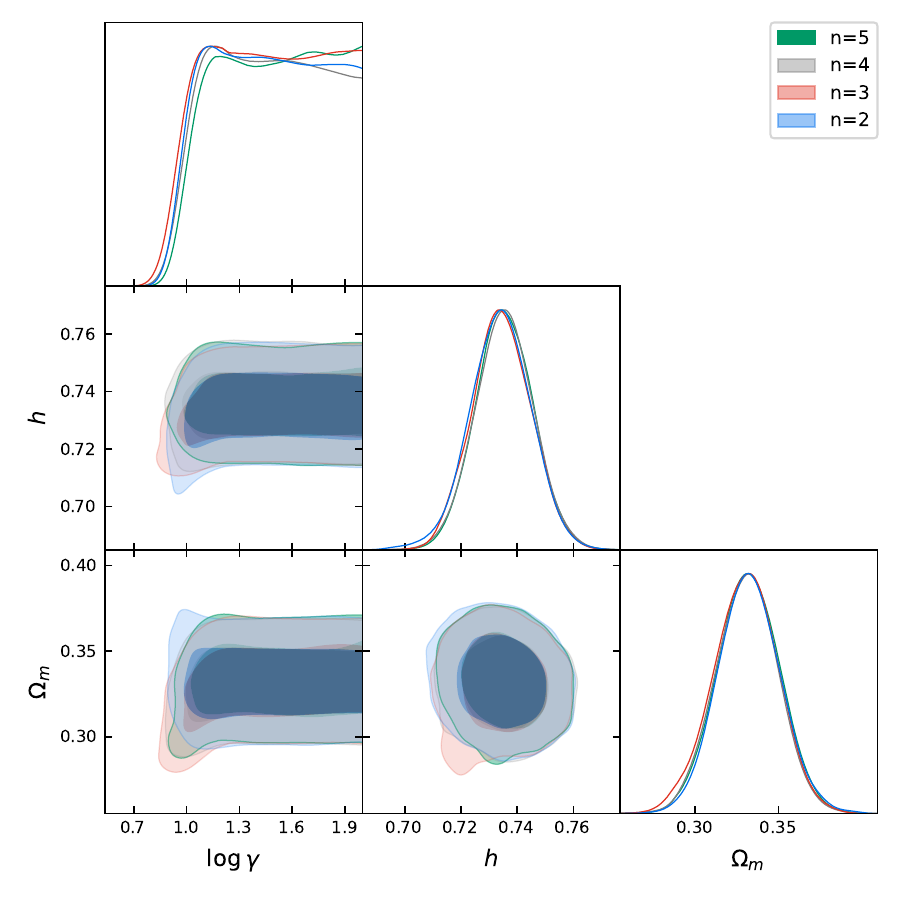}
    \caption{The marginalized 1D and 2D posterior distribution of $\log\gamma$, $h$ and $\Omega_m$ in our model with $n = 5, 4, 3, 2$. We fix $\beta=1/3$ for all models, and we fix $\log\alpha=-2.631, -2.229, -1.838, -1.401$ for  $n = 5,  4,  3,  2$, respectively. Here, we only use the  SNe\,Ia data.}
    \label{fig:MCMC2}
\end{figure}

Figure \ref{fig:EvolutionAll} shows that our model can introduce a component of dynamical dark energy at low $z$. The low-redshift dynamics depend on $\gamma$. Here we present constraints on this parameter based on SH0ES and Pantheon plus data. In this analysis, parameters except $\gamma$, $h$, and $\Omega_m$ are fixed. We assume $\beta=1/3$, $\log\alpha=-2.631, -2.229,-1.838,-1.401$ for  $n = 5, 4, 3, 2$, respectively. Figure \ref{fig:MCMC2} gives the result. As we expected, $\gamma$ must exceed a threshold value for each model under consideration. For the model with fixed $\beta$ and $\alpha$ studied here, we observed at a $95\%$ confidence level that $\log\gamma$ is greater than $1.02$ for $n=5$, $1.00$ for $n=4$, $0.96$ for $n=3$, and $0.98$ for $n=2$. This lower limit result indicates that supernova together with our model does not show any evidence for the dynamical dark energy at low-redshifts.

\begin{figure}[h!]
    \includegraphics[width=0.45\textwidth]{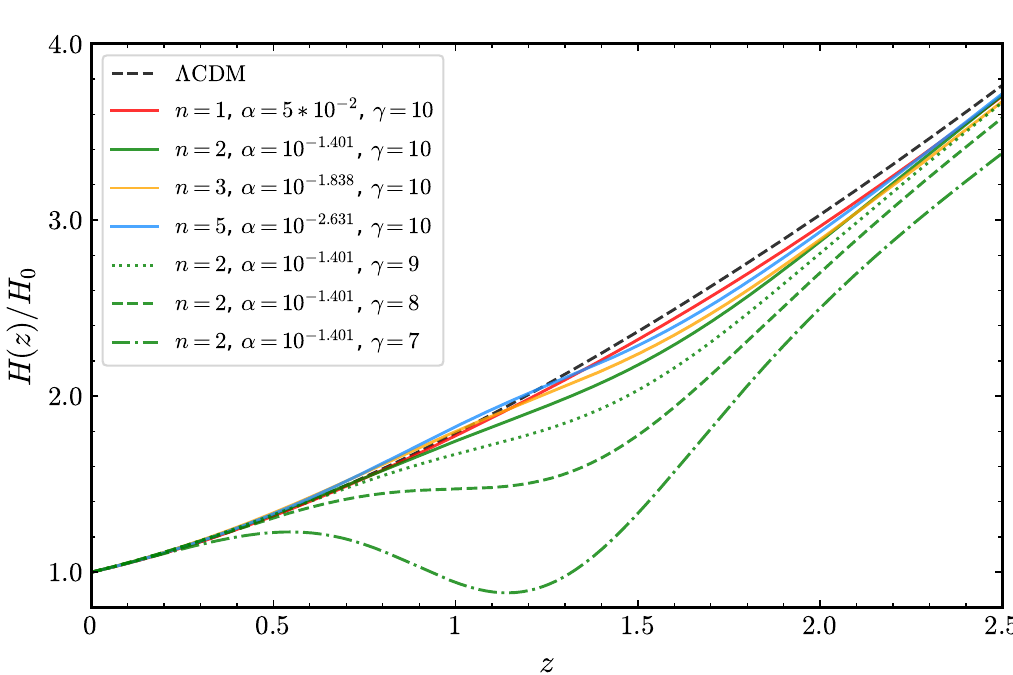}
    \caption{Evolution of $H(z)/H_0$ as a function of redshifts $z$ for EDE models with different parameters. We set $\beta=1/3$, $H_0=73.04$ for each model. }
    \label{fig:evolution_of_H}
\end{figure}

In Fig. \ref{fig:evolution_of_H}, we show the evolution of $H$ for some special parameters. As we discussed in Sec. \ref{sec:Dynamic}, when $\gamma$ is large (here, significantly greater than $10$~\footnote{The base of $\log$ is $10$ in our conventions.}), our model has almost no impact on the late-time evolution of the Universe. However, when the value of $\gamma$ approaches a lower limit, the evolution of the Hubble parameter strongly depends on the value of $\gamma$ and deviates from the $\Lambda$CDM model.

\section{\label{sec:conclution} Discussion and Conclusion}
A novel model designed to address the coincidence problem of EDE was presented in this paper. 
The new EDE theory is based on the fact that we can design geometric quantities that remain zero during the mater and radiation-dominated epoch but nonzero during the radiation-matter transition, as shown in Fig. \ref{fig:f}. We construct such quantities $P$ and use it as a source of the scalar field $\phi$ to realize the EDE. Since this EDE is triggered by the spacetime dynamics during the radiation-matter transition, it can appear naturally near the matter-radiation equality and thus solve the coincidence problem.
	
The new EDE model have $4$ key parameters: $\alpha$, $n$, $\beta$ and $\gamma$. Here, $n$ plays a pivotal role in determining when the EDE becomes active, and larger $n$ corresponds to earlier EDE activation. For models with $n=1$, EDE lags the matter-radiation equality and can persist until low redshift (see Fig. \ref{fig:EvolutionAll}). This case cannot solve the Hubble tension but may provide interesting dynamics in the dark ages of the Universe. For models with $n\geq2$, the peak of EDE appears near matter-radiation equality, which can solve the coincidence problem of EDE. The parameter space with $\{n\geq2,\beta>1/6\}$ can make EDE decay faster than radiation during the matter-dominated era.
	
When dark energy begins to replace dark matter to dominate the Universe, the scalar field $\phi$ again becomes active and leads to an extra energy component. This provides a new theoretical motivation for the dynamical dark energy. However, it should be noted that a large $\gamma$ can suppress this possibility. Constraints from supernova gives a lower limit on $\gamma$.

\begin{acknowledgments}
  This work was supported by the National Natural Science Foundation of China under Grants No. 12021003, No. 11920101003, and No. 11633001, and the Strategic Priority Research Program of the Chinese Academy of Sciences under Grant No. XDB23000000. S. T. was also supported by the Initiative Postdocs Supporting Program under Grant No. BX20200065, China Postdoctoral Science Foundation under Grant No. 2021M700481 and the Fundamental Research Funds for the Central Universities (Beijing Normal University 10800-310400209522).
\end{acknowledgments}


\begin{thebibliography}{47}%
\makeatletter
\providecommand \@ifxundefined [1]{%
 \@ifx{#1\undefined}
}%
\providecommand \@ifnum [1]{%
 \ifnum #1\expandafter \@firstoftwo
 \else \expandafter \@secondoftwo
 \fi
}%
\providecommand \@ifx [1]{%
 \ifx #1\expandafter \@firstoftwo
 \else \expandafter \@secondoftwo
 \fi
}%
\providecommand \natexlab [1]{#1}%
\providecommand \enquote  [1]{``#1''}%
\providecommand \bibnamefont  [1]{#1}%
\providecommand \bibfnamefont [1]{#1}%
\providecommand \citenamefont [1]{#1}%
\providecommand \href@noop [0]{\@secondoftwo}%
\providecommand \href [0]{\begingroup \@sanitize@url \@href}%
\providecommand \@href[1]{\@@startlink{#1}\@@href}%
\providecommand \@@href[1]{\endgroup#1\@@endlink}%
\providecommand \@sanitize@url [0]{\catcode `\\12\catcode `\$12\catcode
  `\&12\catcode `\#12\catcode `\^12\catcode `\_12\catcode `\%12\relax}%
\providecommand \@@startlink[1]{}%
\providecommand \@@endlink[0]{}%
\providecommand \url  [0]{\begingroup\@sanitize@url \@url }%
\providecommand \@url [1]{\endgroup\@href {#1}{\urlprefix }}%
\providecommand \urlprefix  [0]{URL }%
\providecommand \Eprint [0]{\href }%
\providecommand \doibase [0]{https://doi.org/}%
\providecommand \selectlanguage [0]{\@gobble}%
\providecommand \bibinfo  [0]{\@secondoftwo}%
\providecommand \bibfield  [0]{\@secondoftwo}%
\providecommand \translation [1]{[#1]}%
\providecommand \BibitemOpen [0]{}%
\providecommand \bibitemStop [0]{}%
\providecommand \bibitemNoStop [0]{.\EOS\space}%
\providecommand \EOS [0]{\spacefactor3000\relax}%
\providecommand \BibitemShut  [1]{\csname bibitem#1\endcsname}%
\let\auto@bib@innerbib\@empty
\bibitem [{\citenamefont {Aylor}\ \emph {et~al.}(2019)\citenamefont {Aylor},
  \citenamefont {Joy}, \citenamefont {Knox}, \citenamefont {Millea},
  \citenamefont {Raghunathan},\ and\ \citenamefont {Wu}}]{Astrophys.J.874.4}%
  \BibitemOpen
  \bibfield  {author} {\bibinfo {author} {\bibfnamefont {K.}~\bibnamefont
  {Aylor}}, \bibinfo {author} {\bibfnamefont {M.}~\bibnamefont {Joy}}, \bibinfo
  {author} {\bibfnamefont {L.}~\bibnamefont {Knox}}, \bibinfo {author}
  {\bibfnamefont {M.}~\bibnamefont {Millea}}, \bibinfo {author} {\bibfnamefont
  {S.}~\bibnamefont {Raghunathan}},\ and\ \bibinfo {author} {\bibfnamefont
  {W.~L.~K.}\ \bibnamefont {Wu}},\ }\href
  {https://doi.org/10.3847/1538-4357/ab0898} {\bibfield  {journal} {\bibinfo
  {journal} {Astrophys. J.}\ }\textbf {\bibinfo {volume} {874}},\ \bibinfo
  {pages} {4} (\bibinfo {year} {2019})}\BibitemShut {NoStop}%
\bibitem [{\citenamefont {Knox}\ and\ \citenamefont
  {Millea}(2020)}]{PhysRevD.101.043533}%
  \BibitemOpen
  \bibfield  {author} {\bibinfo {author} {\bibfnamefont {L.}~\bibnamefont
  {Knox}}\ and\ \bibinfo {author} {\bibfnamefont {M.}~\bibnamefont {Millea}},\
  }\href {https://doi.org/10.1103/PhysRevD.101.043533} {\bibfield  {journal}
  {\bibinfo  {journal} {Phys. Rev. D}\ }\textbf {\bibinfo {volume} {101}},\
  \bibinfo {pages} {043533} (\bibinfo {year} {2020})}\BibitemShut {NoStop}%
\bibitem [{\citenamefont {{Aghanim, N.}}\ \emph {et~al.}(2020)\citenamefont
  {{Aghanim, N.}}, \citenamefont {{Akrami, Y.}}, \citenamefont {{Ashdown, M.}},
  \citenamefont {{Aumont, J.}}, \citenamefont {{Baccigalupi, C.}},
  \citenamefont {{Ballardini, M.}}, \citenamefont {{Banday, A. J.}},
  \citenamefont {{Barreiro, R. B.}}, \citenamefont {{Bartolo, N.}},
  \citenamefont {{Basak, S.}} \emph {et~al.}}]{Astron.Astrophys.641.A6}%
  \BibitemOpen
  \bibfield  {author} {\bibinfo {author} {\bibnamefont {{Aghanim, N.}}},
  \bibinfo {author} {\bibnamefont {{Akrami, Y.}}}, \bibinfo {author}
  {\bibnamefont {{Ashdown, M.}}}, \bibinfo {author} {\bibnamefont {{Aumont,
  J.}}}, \bibinfo {author} {\bibnamefont {{Baccigalupi, C.}}}, \bibinfo
  {author} {\bibnamefont {{Ballardini, M.}}}, \bibinfo {author} {\bibnamefont
  {{Banday, A. J.}}}, \bibinfo {author} {\bibnamefont {{Barreiro, R. B.}}},
  \bibinfo {author} {\bibnamefont {{Bartolo, N.}}}, \bibinfo {author}
  {\bibnamefont {{Basak, S.}}}, \emph {et~al.} (\bibinfo {collaboration}
  {Planck Collaboration}),\ }\href
  {https://doi.org/10.1051/0004-6361/201833910} {\bibfield  {journal} {\bibinfo
   {journal} {Astron. Astrophys.}\ }\textbf {\bibinfo {volume} {641}},\
  \bibinfo {pages} {A6} (\bibinfo {year} {2020})}\BibitemShut {NoStop}%
\bibitem [{\citenamefont {Riess}\ \emph {et~al.}(2019)\citenamefont {Riess},
  \citenamefont {Casertano}, \citenamefont {Yuan}, \citenamefont {Macri},\ and\
  \citenamefont {Scolnic}}]{Astrophys.J.876.85}%
  \BibitemOpen
  \bibfield  {author} {\bibinfo {author} {\bibfnamefont {A.~G.}\ \bibnamefont
  {Riess}}, \bibinfo {author} {\bibfnamefont {S.}~\bibnamefont {Casertano}},
  \bibinfo {author} {\bibfnamefont {W.}~\bibnamefont {Yuan}}, \bibinfo {author}
  {\bibfnamefont {L.~M.}\ \bibnamefont {Macri}},\ and\ \bibinfo {author}
  {\bibfnamefont {D.}~\bibnamefont {Scolnic}},\ }\href
  {https://doi.org/10.3847/1538-4357/ab1422} {\bibfield  {journal} {\bibinfo
  {journal} {Astrophys. J}\ }\textbf {\bibinfo {volume} {876}},\ \bibinfo
  {pages} {85} (\bibinfo {year} {2019})}\BibitemShut {NoStop}%
\bibitem [{\citenamefont {Riess}\ \emph
  {et~al.}(2022{\natexlab{a}})\citenamefont {Riess}, \citenamefont {Yuan},
  \citenamefont {Macri}, \citenamefont {Scolnic}, \citenamefont {Brout},
  \citenamefont {Casertano}, \citenamefont {Jones}, \citenamefont {Murakami},
  \citenamefont {Anand}, \citenamefont {Breuval} \emph
  {et~al.}}]{Astrophys.J.Lett.934.L7}%
  \BibitemOpen
  \bibfield  {author} {\bibinfo {author} {\bibfnamefont {A.~G.}\ \bibnamefont
  {Riess}}, \bibinfo {author} {\bibfnamefont {W.}~\bibnamefont {Yuan}},
  \bibinfo {author} {\bibfnamefont {L.~M.}\ \bibnamefont {Macri}}, \bibinfo
  {author} {\bibfnamefont {D.}~\bibnamefont {Scolnic}}, \bibinfo {author}
  {\bibfnamefont {D.}~\bibnamefont {Brout}}, \bibinfo {author} {\bibfnamefont
  {S.}~\bibnamefont {Casertano}}, \bibinfo {author} {\bibfnamefont {D.~O.}\
  \bibnamefont {Jones}}, \bibinfo {author} {\bibfnamefont {Y.}~\bibnamefont
  {Murakami}}, \bibinfo {author} {\bibfnamefont {G.~S.}\ \bibnamefont {Anand}},
  \bibinfo {author} {\bibfnamefont {L.}~\bibnamefont {Breuval}}, \emph
  {et~al.},\ }\href {https://doi.org/10.3847/2041-8213/ac5c5b} {\bibfield
  {journal} {\bibinfo  {journal} {Astrophys. J. Lett.}\ }\textbf {\bibinfo
  {volume} {934}},\ \bibinfo {pages} {L7} (\bibinfo {year}
  {2022}{\natexlab{a}})}\BibitemShut {NoStop}%
\bibitem [{\citenamefont {Spergel}\ \emph {et~al.}(2015)\citenamefont
  {Spergel}, \citenamefont {Flauger},\ and\ \citenamefont
  {Hlo\ifmmode~\check{z}\else \v{z}\fi{}ek}}]{PhysRevD.91.023518}%
  \BibitemOpen
  \bibfield  {author} {\bibinfo {author} {\bibfnamefont {D.~N.}\ \bibnamefont
  {Spergel}}, \bibinfo {author} {\bibfnamefont {R.}~\bibnamefont {Flauger}},\
  and\ \bibinfo {author} {\bibfnamefont {R.}~\bibnamefont
  {Hlo\ifmmode~\check{z}\else \v{z}\fi{}ek}},\ }\href
  {https://doi.org/10.1103/PhysRevD.91.023518} {\bibfield  {journal} {\bibinfo
  {journal} {Phys. Rev. D}\ }\textbf {\bibinfo {volume} {91}},\ \bibinfo
  {pages} {023518} (\bibinfo {year} {2015})}\BibitemShut {NoStop}%
\bibitem [{\citenamefont {Rigault}\ \emph {et~al.}(2015)\citenamefont
  {Rigault}, \citenamefont {Aldering}, \citenamefont {Kowalski}, \citenamefont
  {Copin}, \citenamefont {Antilogus}, \citenamefont {Aragon}, \citenamefont
  {Bailey}, \citenamefont {Baltay}, \citenamefont {Baugh}, \citenamefont
  {Bongard} \emph {et~al.}}]{Astrophys.J.802.20}%
  \BibitemOpen
  \bibfield  {author} {\bibinfo {author} {\bibfnamefont {M.}~\bibnamefont
  {Rigault}}, \bibinfo {author} {\bibfnamefont {G.}~\bibnamefont {Aldering}},
  \bibinfo {author} {\bibfnamefont {M.}~\bibnamefont {Kowalski}}, \bibinfo
  {author} {\bibfnamefont {Y.}~\bibnamefont {Copin}}, \bibinfo {author}
  {\bibfnamefont {P.}~\bibnamefont {Antilogus}}, \bibinfo {author}
  {\bibfnamefont {C.}~\bibnamefont {Aragon}}, \bibinfo {author} {\bibfnamefont
  {S.}~\bibnamefont {Bailey}}, \bibinfo {author} {\bibfnamefont
  {C.}~\bibnamefont {Baltay}}, \bibinfo {author} {\bibfnamefont
  {D.}~\bibnamefont {Baugh}}, \bibinfo {author} {\bibfnamefont
  {S.}~\bibnamefont {Bongard}}, \emph {et~al.},\ }\href
  {https://doi.org/10.1088/0004-637X/802/1/20} {\bibfield  {journal} {\bibinfo
  {journal} {Astrophys. J.}\ }\textbf {\bibinfo {volume} {802}},\ \bibinfo
  {pages} {20} (\bibinfo {year} {2015})}\BibitemShut {NoStop}%
\bibitem [{\citenamefont {Addison}\ \emph {et~al.}(2016)\citenamefont
  {Addison}, \citenamefont {Huang}, \citenamefont {Watts}, \citenamefont
  {Bennett}, \citenamefont {Halpern}, \citenamefont {Hinshaw},\ and\
  \citenamefont {Weiland}}]{Astrophys.J.818.132}%
  \BibitemOpen
  \bibfield  {author} {\bibinfo {author} {\bibfnamefont {G.~E.}\ \bibnamefont
  {Addison}}, \bibinfo {author} {\bibfnamefont {Y.}~\bibnamefont {Huang}},
  \bibinfo {author} {\bibfnamefont {D.~J.}\ \bibnamefont {Watts}}, \bibinfo
  {author} {\bibfnamefont {C.~L.}\ \bibnamefont {Bennett}}, \bibinfo {author}
  {\bibfnamefont {M.}~\bibnamefont {Halpern}}, \bibinfo {author} {\bibfnamefont
  {G.}~\bibnamefont {Hinshaw}},\ and\ \bibinfo {author} {\bibfnamefont {J.~L.}\
  \bibnamefont {Weiland}},\ }\href
  {https://doi.org/10.3847/0004-637X/818/2/132} {\bibfield  {journal} {\bibinfo
   {journal} {Astrophys. J.}\ }\textbf {\bibinfo {volume} {818}},\ \bibinfo
  {pages} {132} (\bibinfo {year} {2016})}\BibitemShut {NoStop}%
\bibitem [{\citenamefont {Jones}\ \emph {et~al.}(2018)\citenamefont {Jones},
  \citenamefont {Riess}, \citenamefont {Scolnic}, \citenamefont {Pan},
  \citenamefont {Johnson}, \citenamefont {Coulter}, \citenamefont {Dettman},
  \citenamefont {Foley}, \citenamefont {Foley}, \citenamefont {Huber} \emph
  {et~al.}}]{Astrophys.J.867.108}%
  \BibitemOpen
  \bibfield  {author} {\bibinfo {author} {\bibfnamefont {D.~O.}\ \bibnamefont
  {Jones}}, \bibinfo {author} {\bibfnamefont {A.~G.}\ \bibnamefont {Riess}},
  \bibinfo {author} {\bibfnamefont {D.~M.}\ \bibnamefont {Scolnic}}, \bibinfo
  {author} {\bibfnamefont {Y.-C.}\ \bibnamefont {Pan}}, \bibinfo {author}
  {\bibfnamefont {E.}~\bibnamefont {Johnson}}, \bibinfo {author} {\bibfnamefont
  {D.~A.}\ \bibnamefont {Coulter}}, \bibinfo {author} {\bibfnamefont {K.~G.}\
  \bibnamefont {Dettman}}, \bibinfo {author} {\bibfnamefont {M.~M.}\
  \bibnamefont {Foley}}, \bibinfo {author} {\bibfnamefont {R.~J.}\ \bibnamefont
  {Foley}}, \bibinfo {author} {\bibfnamefont {M.~E.}\ \bibnamefont {Huber}},
  \emph {et~al.},\ }\href {https://doi.org/10.3847/1538-4357/aae2b9} {\bibfield
   {journal} {\bibinfo  {journal} {Astrophys. J.}\ }\textbf {\bibinfo {volume}
  {867}},\ \bibinfo {pages} {108} (\bibinfo {year} {2018})}\BibitemShut
  {NoStop}%
\bibitem [{\citenamefont {Poulin}\ \emph {et~al.}(2019)\citenamefont {Poulin},
  \citenamefont {Smith}, \citenamefont {Karwal},\ and\ \citenamefont
  {Kamionkowski}}]{PhysRevLett.122.221301}%
  \BibitemOpen
  \bibfield  {author} {\bibinfo {author} {\bibfnamefont {V.}~\bibnamefont
  {Poulin}}, \bibinfo {author} {\bibfnamefont {T.~L.}\ \bibnamefont {Smith}},
  \bibinfo {author} {\bibfnamefont {T.}~\bibnamefont {Karwal}},\ and\ \bibinfo
  {author} {\bibfnamefont {M.}~\bibnamefont {Kamionkowski}},\ }\href
  {https://doi.org/10.1103/PhysRevLett.122.221301} {\bibfield  {journal}
  {\bibinfo  {journal} {Phys. Rev. Lett.}\ }\textbf {\bibinfo {volume} {122}},\
  \bibinfo {pages} {221301} (\bibinfo {year} {2019})}\BibitemShut {NoStop}%
\bibitem [{\citenamefont {Braglia}\ \emph {et~al.}(2020)\citenamefont
  {Braglia}, \citenamefont {Emond}, \citenamefont {Finelli}, \citenamefont
  {G\"umr\"uk\ifmmode \mbox{\c{c}}\else \c{c}\fi{}\"uo\ifmmode~\breve{g}\else
  \u{g}\fi{}lu},\ and\ \citenamefont {Koyama}}]{PhysRevD.102.083513}%
  \BibitemOpen
  \bibfield  {author} {\bibinfo {author} {\bibfnamefont {M.}~\bibnamefont
  {Braglia}}, \bibinfo {author} {\bibfnamefont {W.~T.}\ \bibnamefont {Emond}},
  \bibinfo {author} {\bibfnamefont {F.}~\bibnamefont {Finelli}}, \bibinfo
  {author} {\bibfnamefont {A.~E.}\ \bibnamefont {G\"umr\"uk\ifmmode
  \mbox{\c{c}}\else \c{c}\fi{}\"uo\ifmmode~\breve{g}\else \u{g}\fi{}lu}},\ and\
  \bibinfo {author} {\bibfnamefont {K.}~\bibnamefont {Koyama}},\ }\href
  {https://doi.org/10.1103/PhysRevD.102.083513} {\bibfield  {journal} {\bibinfo
   {journal} {Phys. Rev. D}\ }\textbf {\bibinfo {volume} {102}},\ \bibinfo
  {pages} {083513} (\bibinfo {year} {2020})}\BibitemShut {NoStop}%
\bibitem [{\citenamefont {Niedermann}\ and\ \citenamefont
  {Sloth}(2020)}]{PhysRevD.102.063527}%
  \BibitemOpen
  \bibfield  {author} {\bibinfo {author} {\bibfnamefont {F.}~\bibnamefont
  {Niedermann}}\ and\ \bibinfo {author} {\bibfnamefont {M.~S.}\ \bibnamefont
  {Sloth}},\ }\href {https://doi.org/10.1103/PhysRevD.102.063527} {\bibfield
  {journal} {\bibinfo  {journal} {Phys. Rev. D}\ }\textbf {\bibinfo {volume}
  {102}},\ \bibinfo {pages} {063527} (\bibinfo {year} {2020})}\BibitemShut
  {NoStop}%
\bibitem [{\citenamefont {Smith}\ \emph {et~al.}(2020)\citenamefont {Smith},
  \citenamefont {Poulin},\ and\ \citenamefont {Amin}}]{PhysRevD.101.063523}%
  \BibitemOpen
  \bibfield  {author} {\bibinfo {author} {\bibfnamefont {T.~L.}\ \bibnamefont
  {Smith}}, \bibinfo {author} {\bibfnamefont {V.}~\bibnamefont {Poulin}},\ and\
  \bibinfo {author} {\bibfnamefont {M.~A.}\ \bibnamefont {Amin}},\ }\href
  {https://doi.org/10.1103/PhysRevD.101.063523} {\bibfield  {journal} {\bibinfo
   {journal} {Phys. Rev. D}\ }\textbf {\bibinfo {volume} {101}},\ \bibinfo
  {pages} {063523} (\bibinfo {year} {2020})}\BibitemShut {NoStop}%
\bibitem [{\citenamefont {Zumalac\'arregui}(2020)}]{PhysRevD.102.023523}%
  \BibitemOpen
  \bibfield  {author} {\bibinfo {author} {\bibfnamefont {M.}~\bibnamefont
  {Zumalac\'arregui}},\ }\href {https://doi.org/10.1103/PhysRevD.102.023523}
  {\bibfield  {journal} {\bibinfo  {journal} {Phys. Rev. D}\ }\textbf {\bibinfo
  {volume} {102}},\ \bibinfo {pages} {023523} (\bibinfo {year}
  {2020})}\BibitemShut {NoStop}%
\bibitem [{\citenamefont {Braglia}\ \emph {et~al.}(2021)\citenamefont
  {Braglia}, \citenamefont {Ballardini}, \citenamefont {Finelli},\ and\
  \citenamefont {Koyama}}]{PhysRevD.103.043528}%
  \BibitemOpen
  \bibfield  {author} {\bibinfo {author} {\bibfnamefont {M.}~\bibnamefont
  {Braglia}}, \bibinfo {author} {\bibfnamefont {M.}~\bibnamefont {Ballardini}},
  \bibinfo {author} {\bibfnamefont {F.}~\bibnamefont {Finelli}},\ and\ \bibinfo
  {author} {\bibfnamefont {K.}~\bibnamefont {Koyama}},\ }\href
  {https://doi.org/10.1103/PhysRevD.103.043528} {\bibfield  {journal} {\bibinfo
   {journal} {Phys. Rev. D}\ }\textbf {\bibinfo {volume} {103}},\ \bibinfo
  {pages} {043528} (\bibinfo {year} {2021})}\BibitemShut {NoStop}%
\bibitem [{\citenamefont {Chudaykin}\ \emph {et~al.}(2021)\citenamefont
  {Chudaykin}, \citenamefont {Gorbunov},\ and\ \citenamefont
  {Nedelko}}]{PhysRevD.103.043529}%
  \BibitemOpen
  \bibfield  {author} {\bibinfo {author} {\bibfnamefont {A.}~\bibnamefont
  {Chudaykin}}, \bibinfo {author} {\bibfnamefont {D.}~\bibnamefont
  {Gorbunov}},\ and\ \bibinfo {author} {\bibfnamefont {N.}~\bibnamefont
  {Nedelko}},\ }\href {https://doi.org/10.1103/PhysRevD.103.043529} {\bibfield
  {journal} {\bibinfo  {journal} {Phys. Rev. D}\ }\textbf {\bibinfo {volume}
  {103}},\ \bibinfo {pages} {043529} (\bibinfo {year} {2021})}\BibitemShut
  {NoStop}%
\bibitem [{\citenamefont {Valentino}\ \emph {et~al.}(2021)\citenamefont
  {Valentino}, \citenamefont {Mena}, \citenamefont {Pan}, \citenamefont
  {Visinelli}, \citenamefont {Yang}, \citenamefont {Melchiorri}, \citenamefont
  {Mota}, \citenamefont {Riess},\ and\ \citenamefont
  {Silk}}]{DiValentino_2021}%
  \BibitemOpen
  \bibfield  {author} {\bibinfo {author} {\bibfnamefont {E.~D.}\ \bibnamefont
  {Valentino}}, \bibinfo {author} {\bibfnamefont {O.}~\bibnamefont {Mena}},
  \bibinfo {author} {\bibfnamefont {S.}~\bibnamefont {Pan}}, \bibinfo {author}
  {\bibfnamefont {L.}~\bibnamefont {Visinelli}}, \bibinfo {author}
  {\bibfnamefont {W.}~\bibnamefont {Yang}}, \bibinfo {author} {\bibfnamefont
  {A.}~\bibnamefont {Melchiorri}}, \bibinfo {author} {\bibfnamefont {D.~F.}\
  \bibnamefont {Mota}}, \bibinfo {author} {\bibfnamefont {A.~G.}\ \bibnamefont
  {Riess}},\ and\ \bibinfo {author} {\bibfnamefont {J.}~\bibnamefont {Silk}},\
  }\href {https://doi.org/10.1088/1361-6382/ac086d} {\bibfield  {journal}
  {\bibinfo  {journal} {Classical Quantum Gravity}\ }\textbf {\bibinfo {volume}
  {38}},\ \bibinfo {pages} {153001} (\bibinfo {year} {2021})}\BibitemShut
  {NoStop}%
\bibitem [{\citenamefont {Kojima}\ and\ \citenamefont
  {Okubo}(2022)}]{PhysRevD.106.063540}%
  \BibitemOpen
  \bibfield  {author} {\bibinfo {author} {\bibfnamefont {K.}~\bibnamefont
  {Kojima}}\ and\ \bibinfo {author} {\bibfnamefont {Y.}~\bibnamefont {Okubo}},\
  }\href {https://doi.org/10.1103/PhysRevD.106.063540} {\bibfield  {journal}
  {\bibinfo  {journal} {Phys. Rev. D}\ }\textbf {\bibinfo {volume} {106}},\
  \bibinfo {pages} {063540} (\bibinfo {year} {2022})}\BibitemShut {NoStop}%
\bibitem [{\citenamefont {Kamionkowski}\ \emph {et~al.}(2014)\citenamefont
  {Kamionkowski}, \citenamefont {Pradler},\ and\ \citenamefont
  {Walker}}]{PhysRevLett.113.251302}%
  \BibitemOpen
  \bibfield  {author} {\bibinfo {author} {\bibfnamefont {M.}~\bibnamefont
  {Kamionkowski}}, \bibinfo {author} {\bibfnamefont {J.}~\bibnamefont
  {Pradler}},\ and\ \bibinfo {author} {\bibfnamefont {D.~G.~E.}\ \bibnamefont
  {Walker}},\ }\href {https://doi.org/10.1103/PhysRevLett.113.251302}
  {\bibfield  {journal} {\bibinfo  {journal} {Phys. Rev. Lett.}\ }\textbf
  {\bibinfo {volume} {113}},\ \bibinfo {pages} {251302} (\bibinfo {year}
  {2014})}\BibitemShut {NoStop}%
\bibitem [{\citenamefont {Sakstein}\ and\ \citenamefont
  {Trodden}(2020)}]{PhysRevLett.124.161301}%
  \BibitemOpen
  \bibfield  {author} {\bibinfo {author} {\bibfnamefont {J.}~\bibnamefont
  {Sakstein}}\ and\ \bibinfo {author} {\bibfnamefont {M.}~\bibnamefont
  {Trodden}},\ }\href {https://doi.org/10.1103/PhysRevLett.124.161301}
  {\bibfield  {journal} {\bibinfo  {journal} {Phys. Rev. Lett.}\ }\textbf
  {\bibinfo {volume} {124}},\ \bibinfo {pages} {161301} (\bibinfo {year}
  {2020})}\BibitemShut {NoStop}%
\bibitem [{\citenamefont {Gonz\'alez}\ \emph {et~al.}(2021)\citenamefont
  {Gonz\'alez}, \citenamefont {Liang}, \citenamefont {Sakstein},\ and\
  \citenamefont {Trodden}}]{JCAP.2021(06).063}%
  \BibitemOpen
  \bibfield  {author} {\bibinfo {author} {\bibfnamefont {M.~C.}\ \bibnamefont
  {Gonz\'alez}}, \bibinfo {author} {\bibfnamefont {Q.}~\bibnamefont {Liang}},
  \bibinfo {author} {\bibfnamefont {J.}~\bibnamefont {Sakstein}},\ and\
  \bibinfo {author} {\bibfnamefont {M.}~\bibnamefont {Trodden}},\ }\href
  {https://doi.org/10.1088/1475-7516/2021/04/063} {\bibfield  {journal}
  {\bibinfo  {journal} {J. Cosmol. Astropart. Phys.}\ }\bibfield  {number}
  {\bibinfo  {number} {04}} (2021)\ 063}\BibitemShut {NoStop}%
\bibitem [{\citenamefont {Tian}\ and\ \citenamefont
  {Zhu}(2021)}]{PhysRevD.103.043518}%
  \BibitemOpen
  \bibfield  {author} {\bibinfo {author} {\bibfnamefont {S.~X.}\ \bibnamefont
  {Tian}}\ and\ \bibinfo {author} {\bibfnamefont {Z.-H.}\ \bibnamefont {Zhu}},\
  }\href {https://doi.org/10.1103/PhysRevD.103.043518} {\bibfield  {journal}
  {\bibinfo  {journal} {Phys. Rev. D}\ }\textbf {\bibinfo {volume} {103}},\
  \bibinfo {pages} {043518} (\bibinfo {year} {2021})}\BibitemShut {NoStop}%
\bibitem [{\citenamefont {Tian}\ and\ \citenamefont
  {Zhu}(2023)}]{PhysRevD.107.103507}%
  \BibitemOpen
  \bibfield  {author} {\bibinfo {author} {\bibfnamefont {S.~X.}\ \bibnamefont
  {Tian}}\ and\ \bibinfo {author} {\bibfnamefont {Z.-H.}\ \bibnamefont {Zhu}},\
  }\href {https://doi.org/10.1103/PhysRevD.107.103507} {\bibfield  {journal}
  {\bibinfo  {journal} {Phys. Rev. D}\ }\textbf {\bibinfo {volume} {107}},\
  \bibinfo {pages} {103507} (\bibinfo {year} {2023})}\BibitemShut {NoStop}%
\bibitem [{\citenamefont {Karwal}\ \emph {et~al.}(2022)\citenamefont {Karwal},
  \citenamefont {Raveri}, \citenamefont {Jain}, \citenamefont {Khoury},\ and\
  \citenamefont {Trodden}}]{PhysRevD.105.063535}%
  \BibitemOpen
  \bibfield  {author} {\bibinfo {author} {\bibfnamefont {T.}~\bibnamefont
  {Karwal}}, \bibinfo {author} {\bibfnamefont {M.}~\bibnamefont {Raveri}},
  \bibinfo {author} {\bibfnamefont {B.}~\bibnamefont {Jain}}, \bibinfo {author}
  {\bibfnamefont {J.}~\bibnamefont {Khoury}},\ and\ \bibinfo {author}
  {\bibfnamefont {M.}~\bibnamefont {Trodden}},\ }\href
  {https://doi.org/10.1103/PhysRevD.105.063535} {\bibfield  {journal} {\bibinfo
   {journal} {Phys. Rev. D}\ }\textbf {\bibinfo {volume} {105}},\ \bibinfo
  {pages} {063535} (\bibinfo {year} {2022})}\BibitemShut {NoStop}%
\bibitem [{\citenamefont {Lin}\ \emph {et~al.}(2023)\citenamefont {Lin},
  \citenamefont {McDonough}, \citenamefont {Hill},\ and\ \citenamefont
  {Hu}}]{PhysRevD.107.103523}%
  \BibitemOpen
  \bibfield  {author} {\bibinfo {author} {\bibfnamefont {M.-X.}\ \bibnamefont
  {Lin}}, \bibinfo {author} {\bibfnamefont {E.}~\bibnamefont {McDonough}},
  \bibinfo {author} {\bibfnamefont {J.~C.}\ \bibnamefont {Hill}},\ and\
  \bibinfo {author} {\bibfnamefont {W.}~\bibnamefont {Hu}},\ }\href
  {https://doi.org/10.1103/PhysRevD.107.103523} {\bibfield  {journal} {\bibinfo
   {journal} {Phys. Rev. D}\ }\textbf {\bibinfo {volume} {107}},\ \bibinfo
  {pages} {103523} (\bibinfo {year} {2023})}\BibitemShut {NoStop}%
\bibitem [{\citenamefont {Nojiri}\ \emph {et~al.}(2006)\citenamefont {Nojiri},
  \citenamefont {Odintsov},\ and\ \citenamefont {Sami}}]{PhysRevD.74.046004}%
  \BibitemOpen
  \bibfield  {author} {\bibinfo {author} {\bibfnamefont {S.}~\bibnamefont
  {Nojiri}}, \bibinfo {author} {\bibfnamefont {S.~D.}\ \bibnamefont
  {Odintsov}},\ and\ \bibinfo {author} {\bibfnamefont {M.}~\bibnamefont
  {Sami}},\ }\href {https://doi.org/10.1103/PhysRevD.74.046004} {\bibfield
  {journal} {\bibinfo  {journal} {Phys. Rev. D}\ }\textbf {\bibinfo {volume}
  {74}},\ \bibinfo {pages} {046004} (\bibinfo {year} {2006})}\BibitemShut
  {NoStop}%
\bibitem [{\citenamefont {Bartolo}\ and\ \citenamefont
  {Pietroni}(1999)}]{PhysRevD.61.023518}%
  \BibitemOpen
  \bibfield  {author} {\bibinfo {author} {\bibfnamefont {N.}~\bibnamefont
  {Bartolo}}\ and\ \bibinfo {author} {\bibfnamefont {M.}~\bibnamefont
  {Pietroni}},\ }\href {https://doi.org/10.1103/PhysRevD.61.023518} {\bibfield
  {journal} {\bibinfo  {journal} {Phys. Rev. D}\ }\textbf {\bibinfo {volume}
  {61}},\ \bibinfo {pages} {023518} (\bibinfo {year} {1999})}\BibitemShut
  {NoStop}%
\bibitem [{\citenamefont {Esposito-Far\`ese}\ and\ \citenamefont
  {Polarski}(2001)}]{PhysRevD.63.063504}%
  \BibitemOpen
  \bibfield  {author} {\bibinfo {author} {\bibfnamefont {G.}~\bibnamefont
  {Esposito-Far\`ese}}\ and\ \bibinfo {author} {\bibfnamefont {D.}~\bibnamefont
  {Polarski}},\ }\href {https://doi.org/10.1103/PhysRevD.63.063504} {\bibfield
  {journal} {\bibinfo  {journal} {Phys. Rev. D}\ }\textbf {\bibinfo {volume}
  {63}},\ \bibinfo {pages} {063504} (\bibinfo {year} {2001})}\BibitemShut
  {NoStop}%
\bibitem [{\citenamefont {Amendola}\ and\ \citenamefont
  {Tsujikawa}(2010)}]{amendolatsujikawa2010}%
  \BibitemOpen
  \bibfield  {author} {\bibinfo {author} {\bibfnamefont {L.}~\bibnamefont
  {Amendola}}\ and\ \bibinfo {author} {\bibfnamefont {S.}~\bibnamefont
  {Tsujikawa}},\ }\href {https://doi.org/10.1017/CBO9780511750823.010} {\emph
  {\bibinfo {title} {Dark Energy: Theory and Observations}}}\ (\bibinfo
  {publisher} {Cambridge University Press},\ \bibinfo {address} {Cambridge},\
  \bibinfo {year} {2010})\ pp.\ \bibinfo {pages} {234--284}\BibitemShut
  {NoStop}%
\bibitem [{Note1()}]{Note1}%
  \BibitemOpen
  \bibinfo {note} {In general relativity, $R=0$ in the radiation-dominated era,
  and $R\protect \neq 0$ in the matter-dominated era.}\BibitemShut {Stop}%
\bibitem [{\citenamefont {Turner}(1983)}]{PhysRevD.28.1243}%
  \BibitemOpen
  \bibfield  {author} {\bibinfo {author} {\bibfnamefont {M.~S.}\ \bibnamefont
  {Turner}},\ }\href {https://doi.org/10.1103/PhysRevD.28.1243} {\bibfield
  {journal} {\bibinfo  {journal} {Phys. Rev. D}\ }\textbf {\bibinfo {volume}
  {28}},\ \bibinfo {pages} {1243} (\bibinfo {year} {1983})}\BibitemShut
  {NoStop}%
\bibitem [{Note2()}]{Note2}%
  \BibitemOpen
  \bibinfo {note} {\protect \url {http://xact.es}}\BibitemShut {NoStop}%
\bibitem [{Note3()}]{Note3}%
  \BibitemOpen
  \bibinfo {note} {The details are integrated into a public \protect \texttt
  {Mathematica} code, which is available at GitHub (\protect \url
  {https://github.com/JingChang-cheng/MG}).}\BibitemShut {Stop}%
\bibitem [{\citenamefont {Karwal}\ and\ \citenamefont
  {Kamionkowski}(2016)}]{PhysRevD.94.103523}%
  \BibitemOpen
  \bibfield  {author} {\bibinfo {author} {\bibfnamefont {T.}~\bibnamefont
  {Karwal}}\ and\ \bibinfo {author} {\bibfnamefont {M.}~\bibnamefont
  {Kamionkowski}},\ }\href {https://doi.org/10.1103/PhysRevD.94.103523}
  {\bibfield  {journal} {\bibinfo  {journal} {Phys. Rev. D}\ }\textbf {\bibinfo
  {volume} {94}},\ \bibinfo {pages} {103523} (\bibinfo {year}
  {2016})}\BibitemShut {NoStop}%
\bibitem [{\citenamefont {Alexander}\ and\ \citenamefont
  {McDonough}(2019)}]{ALEXANDER2019134830}%
  \BibitemOpen
  \bibfield  {author} {\bibinfo {author} {\bibfnamefont {S.}~\bibnamefont
  {Alexander}}\ and\ \bibinfo {author} {\bibfnamefont {E.}~\bibnamefont
  {McDonough}},\ }\href
  {https://doi.org/https://doi.org/10.1016/j.physletb.2019.134830} {\bibfield
  {journal} {\bibinfo  {journal} {Phys. Lett. B}\ }\textbf {\bibinfo {volume}
  {797}},\ \bibinfo {pages} {134830} (\bibinfo {year} {2019})}\BibitemShut
  {NoStop}%
\bibitem [{\citenamefont {Landau}\ and\ \citenamefont
  {Lifshitz}(1976)}]{Landanvolume1}%
  \BibitemOpen
  \bibfield  {author} {\bibinfo {author} {\bibfnamefont {L.~D.}\ \bibnamefont
  {Landau}}\ and\ \bibinfo {author} {\bibfnamefont {E.~M.}\ \bibnamefont
  {Lifshitz}},\ }\href@noop {} {\emph {\bibinfo {title} {Mechanics}}}\
  (\bibinfo  {publisher} {Butterworth-Heinemann},\ \bibinfo {address}
  {Oxford},\ \bibinfo {year} {1976})\ pp.\ \bibinfo {pages}
  {74--77}\BibitemShut {NoStop}%
\bibitem [{\citenamefont {{Zhao}}\ \emph {et~al.}(2017)\citenamefont {{Zhao}},
  \citenamefont {{Raveri}}, \citenamefont {{Pogosian}}, \citenamefont {{Wang}},
  \citenamefont {{Crittenden}}, \citenamefont {{Handley}}, \citenamefont
  {{Percival}}, \citenamefont {{Beutler}}, \citenamefont {{Brinkmann}},
  \citenamefont {{Chuang}} \emph {et~al.}}]{Zhao2017.NatAstron.1.627}%
  \BibitemOpen
  \bibfield  {author} {\bibinfo {author} {\bibfnamefont {G.-B.}\ \bibnamefont
  {{Zhao}}}, \bibinfo {author} {\bibfnamefont {M.}~\bibnamefont {{Raveri}}},
  \bibinfo {author} {\bibfnamefont {L.}~\bibnamefont {{Pogosian}}}, \bibinfo
  {author} {\bibfnamefont {Y.}~\bibnamefont {{Wang}}}, \bibinfo {author}
  {\bibfnamefont {R.~G.}\ \bibnamefont {{Crittenden}}}, \bibinfo {author}
  {\bibfnamefont {W.~J.}\ \bibnamefont {{Handley}}}, \bibinfo {author}
  {\bibfnamefont {W.~J.}\ \bibnamefont {{Percival}}}, \bibinfo {author}
  {\bibfnamefont {F.}~\bibnamefont {{Beutler}}}, \bibinfo {author}
  {\bibfnamefont {J.}~\bibnamefont {{Brinkmann}}}, \bibinfo {author}
  {\bibfnamefont {C.-H.}\ \bibnamefont {{Chuang}}}, \emph {et~al.},\ }\href
  {https://doi.org/10.1038/s41550-017-0216-z} {\bibfield  {journal} {\bibinfo
  {journal} {Nat. Astron.}\ }\textbf {\bibinfo {volume} {1}},\ \bibinfo {pages}
  {627} (\bibinfo {year} {2017})}\BibitemShut {NoStop}%
\bibitem [{\citenamefont {{Wang}}(2022)}]{Wang2022.PRD.106.063515}%
  \BibitemOpen
  \bibfield  {author} {\bibinfo {author} {\bibfnamefont {D.}~\bibnamefont
  {{Wang}}},\ }\href {https://doi.org/10.1103/PhysRevD.106.063515} {\bibfield
  {journal} {\bibinfo  {journal} {Phys. Rev. D}\ }\textbf {\bibinfo {volume}
  {106}},\ \bibinfo {eid} {063515} (\bibinfo {year} {2022})}\BibitemShut
  {NoStop}%
\bibitem [{\citenamefont {Wetterich}(2004)}]{PLB.594.17.2004}%
  \BibitemOpen
  \bibfield  {author} {\bibinfo {author} {\bibfnamefont {C.}~\bibnamefont
  {Wetterich}},\ }\href
  {https://doi.org/https://doi.org/10.1016/j.physletb.2004.05.008} {\bibfield
  {journal} {\bibinfo  {journal} {Phys. Lett. B}\ }\textbf {\bibinfo {volume}
  {594}},\ \bibinfo {pages} {17} (\bibinfo {year} {2004})}\BibitemShut
  {NoStop}%
\bibitem [{\citenamefont {Pettorino}\ \emph {et~al.}(2013)\citenamefont
  {Pettorino}, \citenamefont {Amendola},\ and\ \citenamefont
  {Wetterich}}]{PhysRevD.87.083009}%
  \BibitemOpen
  \bibfield  {author} {\bibinfo {author} {\bibfnamefont {V.}~\bibnamefont
  {Pettorino}}, \bibinfo {author} {\bibfnamefont {L.}~\bibnamefont
  {Amendola}},\ and\ \bibinfo {author} {\bibfnamefont {C.}~\bibnamefont
  {Wetterich}},\ }\href {https://doi.org/10.1103/PhysRevD.87.083009} {\bibfield
   {journal} {\bibinfo  {journal} {Phys. Rev. D}\ }\textbf {\bibinfo {volume}
  {87}},\ \bibinfo {pages} {083009} (\bibinfo {year} {2013})}\BibitemShut
  {NoStop}%
\bibitem [{\citenamefont {Wang}\ and\ \citenamefont
  {Mukherjee}(2007)}]{PhysRevD.76.103533}%
  \BibitemOpen
  \bibfield  {author} {\bibinfo {author} {\bibfnamefont {Y.}~\bibnamefont
  {Wang}}\ and\ \bibinfo {author} {\bibfnamefont {P.}~\bibnamefont
  {Mukherjee}},\ }\href {https://doi.org/10.1103/PhysRevD.76.103533} {\bibfield
   {journal} {\bibinfo  {journal} {Phys. Rev. D}\ }\textbf {\bibinfo {volume}
  {76}},\ \bibinfo {pages} {103533} (\bibinfo {year} {2007})}\BibitemShut
  {NoStop}%
\bibitem [{\citenamefont {Wang}\ and\ \citenamefont
  {Wang}(2013)}]{PhysRevD.88.043522}%
  \BibitemOpen
  \bibfield  {author} {\bibinfo {author} {\bibfnamefont {Y.}~\bibnamefont
  {Wang}}\ and\ \bibinfo {author} {\bibfnamefont {S.}~\bibnamefont {Wang}},\
  }\href {https://doi.org/10.1103/PhysRevD.88.043522} {\bibfield  {journal}
  {\bibinfo  {journal} {Phys. Rev. D}\ }\textbf {\bibinfo {volume} {88}},\
  \bibinfo {pages} {043522} (\bibinfo {year} {2013})}\BibitemShut {NoStop}%
\bibitem [{\citenamefont {Zhai}\ and\ \citenamefont
  {Wang}(2019)}]{JCAP07(2019)005}%
  \BibitemOpen
  \bibfield  {author} {\bibinfo {author} {\bibfnamefont {Z.}~\bibnamefont
  {Zhai}}\ and\ \bibinfo {author} {\bibfnamefont {Y.}~\bibnamefont {Wang}},\
  }\href {https://doi.org/10.1088/1475-7516/2019/07/005} {\bibfield  {journal}
  {\bibinfo  {journal} {J. Cosmol. Astropart. Phys.}\ }\bibfield  {number}
  {\bibinfo  {number} {07}} (2019)\ 005}\BibitemShut {NoStop}%
\bibitem [{\citenamefont {Riess}\ \emph
  {et~al.}(2022{\natexlab{b}})\citenamefont {Riess}, \citenamefont {Yuan},
  \citenamefont {Macri}, \citenamefont {Scolnic}, \citenamefont {Brout},
  \citenamefont {Casertano}, \citenamefont {Jones}, \citenamefont {Murakami},
  \citenamefont {Anand}, \citenamefont {Breuval} \emph
  {et~al.}}]{APJL.934.L7(2022)}%
  \BibitemOpen
  \bibfield  {author} {\bibinfo {author} {\bibfnamefont {A.~G.}\ \bibnamefont
  {Riess}}, \bibinfo {author} {\bibfnamefont {W.}~\bibnamefont {Yuan}},
  \bibinfo {author} {\bibfnamefont {L.~M.}\ \bibnamefont {Macri}}, \bibinfo
  {author} {\bibfnamefont {D.}~\bibnamefont {Scolnic}}, \bibinfo {author}
  {\bibfnamefont {D.}~\bibnamefont {Brout}}, \bibinfo {author} {\bibfnamefont
  {S.}~\bibnamefont {Casertano}}, \bibinfo {author} {\bibfnamefont {D.~O.}\
  \bibnamefont {Jones}}, \bibinfo {author} {\bibfnamefont {Y.}~\bibnamefont
  {Murakami}}, \bibinfo {author} {\bibfnamefont {G.~S.}\ \bibnamefont {Anand}},
  \bibinfo {author} {\bibfnamefont {L.}~\bibnamefont {Breuval}}, \emph
  {et~al.},\ }\href {https://doi.org/10.3847/2041-8213/ac5c5b} {\bibfield
  {journal} {\bibinfo  {journal} {Astrophys. J. Lett.}\ }\textbf {\bibinfo
  {volume} {934}},\ \bibinfo {pages} {L7} (\bibinfo {year}
  {2022}{\natexlab{b}})}\BibitemShut {NoStop}%
\bibitem [{\citenamefont {Brout}\ \emph {et~al.}(2022)\citenamefont {Brout},
  \citenamefont {Scolnic}, \citenamefont {Popovic}, \citenamefont {Riess},
  \citenamefont {Carr}, \citenamefont {Zuntz}, \citenamefont {Kessler},
  \citenamefont {Davis}, \citenamefont {Hinton}, \citenamefont {Jones} \emph
  {et~al.}}]{APJ.938.110(2022)}%
  \BibitemOpen
  \bibfield  {author} {\bibinfo {author} {\bibfnamefont {D.}~\bibnamefont
  {Brout}}, \bibinfo {author} {\bibfnamefont {D.}~\bibnamefont {Scolnic}},
  \bibinfo {author} {\bibfnamefont {B.}~\bibnamefont {Popovic}}, \bibinfo
  {author} {\bibfnamefont {A.~G.}\ \bibnamefont {Riess}}, \bibinfo {author}
  {\bibfnamefont {A.}~\bibnamefont {Carr}}, \bibinfo {author} {\bibfnamefont
  {J.}~\bibnamefont {Zuntz}}, \bibinfo {author} {\bibfnamefont
  {R.}~\bibnamefont {Kessler}}, \bibinfo {author} {\bibfnamefont {T.~M.}\
  \bibnamefont {Davis}}, \bibinfo {author} {\bibfnamefont {S.}~\bibnamefont
  {Hinton}}, \bibinfo {author} {\bibfnamefont {D.}~\bibnamefont {Jones}}, \emph
  {et~al.},\ }\href {https://doi.org/10.3847/1538-4357/ac8e04} {\bibfield
  {journal} {\bibinfo  {journal} {Astrophys. J.}\ }\textbf {\bibinfo {volume}
  {938}},\ \bibinfo {pages} {110} (\bibinfo {year} {2022})}\BibitemShut
  {NoStop}%
\bibitem [{\citenamefont {Zuntz}\ \emph {et~al.}(2015)\citenamefont {Zuntz},
  \citenamefont {Paterno}, \citenamefont {Jennings}, \citenamefont {Rudd},
  \citenamefont {Manzotti}, \citenamefont {Dodelson}, \citenamefont {Bridle},
  \citenamefont {Sehrish},\ and\ \citenamefont {Kowalkowski}}]{ZUNTZ201545}%
  \BibitemOpen
  \bibfield  {author} {\bibinfo {author} {\bibfnamefont {J.}~\bibnamefont
  {Zuntz}}, \bibinfo {author} {\bibfnamefont {M.}~\bibnamefont {Paterno}},
  \bibinfo {author} {\bibfnamefont {E.}~\bibnamefont {Jennings}}, \bibinfo
  {author} {\bibfnamefont {D.}~\bibnamefont {Rudd}}, \bibinfo {author}
  {\bibfnamefont {A.}~\bibnamefont {Manzotti}}, \bibinfo {author}
  {\bibfnamefont {S.}~\bibnamefont {Dodelson}}, \bibinfo {author}
  {\bibfnamefont {S.}~\bibnamefont {Bridle}}, \bibinfo {author} {\bibfnamefont
  {S.}~\bibnamefont {Sehrish}},\ and\ \bibinfo {author} {\bibfnamefont
  {J.}~\bibnamefont {Kowalkowski}},\ }\href
  {https://doi.org/https://doi.org/10.1016/j.ascom.2015.05.005} {\bibfield
  {journal} {\bibinfo  {journal} {Astronomy and Computing}\ }\textbf {\bibinfo
  {volume} {12}},\ \bibinfo {pages} {45} (\bibinfo {year} {2015})}\BibitemShut
  {NoStop}%
\bibitem [{Note4()}]{Note4}%
  \BibitemOpen
  \bibinfo {note} {The base of $\protect \qopname \relax o{log}$ is $10$ in our
  conventions.}\BibitemShut {Stop}%
\end{thebibliography}
\end{document}